\documentclass[a4paper,10pt,twoside]{cpc-hepnp}
\usepackage{CJK,upgreek,fancyhdr}
\usepackage{multicol}
\usepackage{graphicx}
\usepackage{booktabs}
\usepackage{amssymb,bm,mathrsfs,bbm,amscd}
\usepackage[tbtags]{amsmath}
\usepackage{lastpage}
\usepackage{longtable}
\usepackage{changes}
\usepackage{lineno,hyperref}
\usepackage{threeparttable} 
\usepackage{multirow}
\usepackage{subfigure} 
\usepackage{ulem}
\begin{document}
\begin{CJK*}{GB}{gbsn}

\fancyhead[c]{\small Chinese Physics C~~~Vol. xx, No. x (201x) xxxxxx}
\fancyfoot[C]{\small 010201-\thepage}


\title{Systematic study of the $\alpha$ decay preformation factor of nuclei around the $\boldsymbol{Z=82}$, $\boldsymbol{N=126}$ shell closures within a generalized liquid drop model
\thanks{We would like to thank X. -D. Sun, J. -G. Deng, J. -H. Cheng, and J. -L. Chen for useful discussion. This work is supported in part by the National Natural Science Foundation of China (Grants No. 11205083, No.11505100 and No. 11705055), the Construct Program of the Key Discipline in Hunan Province, the Research Foundation of Education Bureau of Hunan Province, China (Grant No. 15A159 and No.18A237), the Natural Science Foundation of Hunan Province, China (Grants No. 2015JJ3103, No. 2015JJ2121 and No. 2018JJ3324), the Innovation Group of Nuclear and Particle Physics in USC, the Shandong Province Natural Science Foundation, China (Grant No. ZR2015AQ007), the Hunan Provincial Innovation Foundation For Postgraduate (Grant No. CX20190714), the National Innovation Training Foundation of China (Grant No.201910555161), and the Opening Project of Cooperative Innovation Center for Nuclear Fuel Cycle Technology and Equipment, University of South China (Grant No. 2019KFZ10).}}
\author{%
    Hong-Ming Liu (刘宏铭)$^{1}$
\quad You-Tian Zou (邹有甜)$^{1}$
\quad Xiao Pan (潘霄)$^{1}$\\
\quad Xiao-Jun Bao (包小军)$^{2,4;1)}$\email{{baoxiaojun@hunnu.edu.cn}}
\quad Xiao-Hua Li (李小华)$^{1,3,4;2)}$\email{lixiaohuaphysics@126.com}%
}
\maketitle

\address{%
$^1$ School of Nuclear Science and Technology, University of South China, Hengyang 421001, China\\
$^2$ Hunan Normal University, Changsha 410081, People's Republic of China\\
$^3$ Cooperative Innovation Center for Nuclear Fuel Cycle Technology $\&$ Equipment, University of South China, Hengyang 421001, China\\
$^4$ Key Laboratory of Low Dimensional Quantum Structures and Quantum Control, Hunan Normal University, Changsha 410081, China\\
}

\begin{abstract}
In this work, we systematically study the $\alpha$ decay preformation factors $P_{\alpha}$ and $\alpha$ decay half-lives of 152 nuclei around $\emph{Z}$ = 82, $\emph{N}$ = 126 closed shells based on a generalized liquid drop model while $P_{\alpha}$ is extracted from the ratio of the calculated $\alpha$ decay half-life to the experimental one. The results show that there is an obvious linear relationship between $P_{\alpha}$ and the product of valance protons (holes) $N_p$ and valance neutrons (holes) $N_n$. At the same time, we extract the $\alpha$ decay preformation factors values of even-even nuclei around $\emph{Z}$ = 82, $\emph{N}$ = 126 closed shells from the work of Sun ${et\ al.}$ [\href {https://doi.org/10.1088/1361-6471/aac981} {J. Phys. G: Nucl. Part. Phys. $\bm{45}$, 075106 (2018)}], in which the $\alpha$ decay preformation factors can be calculated by two different microscopic formulas. We find that the $\alpha$ decay preformation factors are also related to $N_pN_n$. Combining with our previous works [Sun ${et\ al.}$, \href {https://doi.org/10.1103/PhysRevC.94.024338} {Phys. Rev. C $\bm{94}$, 024338 (2016)}; Deng ${et\ al.}$, \href {https://doi.org/10.1103/ PhysRevC 96.024318} {ibid. $\bm{96}$, 024318 (2017)}; Deng ${et\ al.}$, \href {https://doi.org/10.1103/PhysRevC.97.044322} {ibid. $\bm{97}$, 044322 (2018)}] and the work of Seif ${et\ al.}$ [\href {http://dx.doi.org/10.1103/PhysRevC.84.064608}{Phys. Rev. C $\bm{84}$, 064608 (2011)}], we suspect that this phenomenon of linear relationship for the nuclei around those closed shells is model independent. It may be caused by the effect of the valence protons (holes) and valence neutrons (holes) around the shell closures. Finally, using the formula obtained by fitting the $\alpha$ decay preformation factor data calculated by the generalized liquid drop model (GLDM), we calculate the $\alpha$ decay half-lives of these nuclei. The calculated results are agree with the experimental data well.
\end{abstract}



\footnotetext[0]{\hspace*{-3mm}\raisebox{0.3ex}{$\scriptstyle\copyright$}2020
Chinese Physical Society and the Institute of High Energy Physics
of the Chinese Academy of Sciences and the Institute
of Modern Physics of the Chinese Academy of Sciences and IOP Publishing Ltd}%

\begin{multicols}{2}

\section{Introduction}

$\alpha$ decay, which is one of the most significant tools for exploring nuclear structure information, can provide the information of ground-state lifetime, nuclear force, nuclear matter incompressibility, spin and parity of nuclei \cite{PhysRevC.73.014315,PhysRevC.73.014612,PhysRevC.74.034302,Basu_2004}. In 1928, Gamow, Condon and Gurney independently proposed the quantum tunneling theory\cite{Gur28,Gamow1928} named Gamow theory. Within this theory, the $\alpha$ decay is explained as an $\alpha$ cluster preformed in the surface of the parent nucleus penetrating the Coulomb barrier between the cluster and the daughter nucleus. The probability of $\alpha$ cluster formation in the parent nucleus is described as the $\alpha$ preformation factor $P_{\alpha}$, including many nuclear structure information. It has became a hot topic in the nuclear physics \cite{QI201677,PhysRevC.77.054318,PhysRevC.92.044302,PhysRevC.95.061306}.

Up to now, there are many microscopical and phenomenological models have been used to calculate $P_{\alpha}$ \cite{PhysRevLett.69.37,DODIGCRNKOVIC1985419,TONOZUKA197945,DODIGCRNKOVIC1989533,VARGA1992421, PhysRevC.90.034304, PhysRevC.93.011306,XU2005303,Saleh_Ahmed_2013,Deng_2015,PhysRevC.93.044326,SALEHAHMED2017103,PhysRevC.84.027303,PhysRevC.80.064325,Qian2013}. Microscopically, in the \emph{R}-matrix method\cite{PhysRevLett.69.37,DODIGCRNKOVIC1985419,TONOZUKA197945,DODIGCRNKOVIC1989533,VARGA1992421}, $P_{\alpha}$ can be obtained by the initial tailored wavefunction of the parent nucleus. However, the calculation of purely microcosmic $P_{\alpha}$ is very difficult due to the complexity of the nuclear many-body problem and the uncertainty of the nuclear potential. The $\alpha$ preformation factor can also obtained by the approach of the microscopical Tohsaki-Horiuchi-Schuck-R$\ddot{\rm{o}}$pke wave function \cite{PhysRevC.90.034304, PhysRevC.93.011306}, which has been successfully emploied to descibe the cluster structure in light nuclei\cite{PhysRevC.93.011306}. Using the cluster-configuration shell model, Varga ${et\ al.}$ reproduced the experimental decay width of $^{212}$Po and obtained its $\alpha$ preformation factor $P_{\alpha}$ = 0.23\cite{PhysRevLett.69.37,VARGA1992421}. Recently, Ahmed ${et\ al.}$ proposed a new quantum-mechanical theory named cluster-formation model (CFM) to calculate the $\alpha$ preformation factors $P_{\alpha}$ of even-even nuclei \cite{Saleh_Ahmed_2013, Saleh Ahmed2013}. Within this model, $P_{\alpha}$ can be obtained through the ratio of the formation energy $E_{f\alpha}$ to the total energy $E$, while the $E_{f\alpha}$ and $E$ can be obtained by binding energies of the nucleus and its neighboring nuclides. Later, Deng ${et\ al.}$ and Ahmed ${et\ al.}$ extended this model to odd-$\emph{A}$ and odd-odd nuclei\cite{Deng_2015,PhysRevC.93.044326,SALEHAHMED2017103}. Generally, within different theoretical models, the $\alpha$ preformation factor $P_{\alpha}$ was different due to the penetration probability varies greatly with the exponential factor. Phenomenologically, the  preformation factors $P_{\alpha}$ are extracted from the ratios of calculated $\alpha$ decay half-lives to the experimental ones \cite{PhysRevC.84.027303,PhysRevC.80.064325,Qian2013}. In 2005, using the density-dependent cluster model (DDCM)\cite{XU2005303}, Xu and Ren studied the available experimental $\alpha$ decay half-lives of the medium mass nuclei. Their results showed that the $P_{\alpha}$ are different for the different types of parent nuclei i.e. $P_{\alpha}$ = 0.43 for even-even nuclei, 0.35 for odd-$\emph{A}$ nuclei, and 0.18 for doubly odd nuclei.

Recent works have been shown that the $\alpha$ preformation factors are affected by many confirmed factors such as the isospin asymmetry of the parent nucleus, the deformation of the daughter, the pairing effect and the shell effect \cite{Seif2015,Seif2013}. It has been observed that the minima of the $\alpha$ preformation factors are at the protons, neutron shells and subshell closures\cite{Seif2015, PhysRevC.77. 054318, PhysRevLett.110.242502, PhysRevC.89.034617, PhysRevC.100.044302, ProgPartNuclPhys.105.214}. Moreover, many nuclear quantities \cite{PhysRevLett.70.402,PhysRevC.63.067302,PhysRevC.46.R1587,PhysRevLett.94.202501,PhysRevC.33.1819} such as deformation and B (E2) values \cite{PhysRevLett.70.402,PhysRevC.63.067302}, rotational moments of inertia in low spin states in the rare earth region \cite{PhysRevC.46.R1587}, core cluster decomposition in the rare earth region \cite{PhysRevLett.94.202501}, properties of excited states \cite{PhysRevC.33.1819} and so on display a systematic behavior with the product of valence protons (holes) $N_p$ and valence neutrons (holes) $N_n$ of the parent nucleus, respectively. In 2011, Seif \emph{et\ al.} found that the $\alpha$ preformation factor is linearly proportional to the product of the $N_p$ and $N_n$ for even-even nuclei around  $Z$=82 and $N$=126 closed shells \cite{PhysRevC.84.064608}. In our previous works \cite{PhysRevC.94.024338,PhysRevC.96.024318,PhysRevC.97.044322}, based on the two-potential approach (TPA) \cite{PhysRevLett.59.262,PhysRevA.69.042705}, we found that this linear relationship also exist in the cases of odd-$\emph{A}$ and doubly-odd nuclei for favored and unfavored $\alpha$ decay\cite{PhysRevC.94.024338,PhysRevC.96.024318}. Very recently, using cluster-formation model (CFM), we calculated the $\alpha$ preformation factor of all kinds of nuclei and found that this linear relationship is also exist \cite{PhysRevC.97.044322}. Combined with Seif \emph{et\ al.} work and our previous works, it is interesting to validate whether this linear relationship is model dependent or due to the valence proton-neutron interaction of the shell closures. In present work, using  generalized liquid drop model (GLDM) \cite{BLOCKI1977427,Royer_2000,ROYER2001182,PhysRevC.74.017304,ROYER1985477,Bao_2012,BAO20131}, we systematically study the $\alpha$ preformation factors $P_{\alpha}$ and $\alpha$ decay half-lives of 152 nuclei around $\emph{Z}$=82, $\emph{N}$=126 shell closures. Our results show that the $\alpha$ preformation factors of these nuclei and $N_p$$N_n$ still satisfy this linear relationship, which means that this relationship may be caused by the valence proton-neutron correlation around $\emph{Z}$ = 82, $\emph{N}$ = 126 shell closures.
   
The article is arranged as follows. In the next section, the theoretical framework of the GLDM is briefly presented. The detailed calculations and discussion are presented in Section 3. Finally, a summary is given in Section 4.

\section{Theoretical framework}
\label{section 2}

The $\alpha$ decay half-life can be calaulated by decay constant $\lambda$, which can be written as
\begin{equation}
T_{1/2} = \frac{ ln2}{\lambda}.
\end{equation}
Here the $\alpha$ decay constant is defined as
\begin{equation}
\lambda = P_{\alpha} \nu \emph{P},
\end{equation}
where $P_{\alpha}$ is the $\alpha$ preformation factor. $P$, the penetration probability of the $\alpha$ particle crossing the barrier, is calculated by Eq.\,(\ref{subeq:11}). $\nu$ is the assault frequency which can be calculated with the oscillation frequency $\omega$ and written as \cite{PhysRevC.81.064309}
\begin{equation}
\nu = \frac{\omega}{2 \pi} = \frac{(2 n_r + l + \frac{3}{2}) \hbar}{2 \pi \mu {R_n}^{2}}=\frac{(G + \frac{3}{2}) \hbar}{1.2 \pi \mu {R_{00}}^{2}},
\end{equation}
where $\mu$ = $\frac{m_d m_\alpha}{m_d + m_\alpha}$ represents the reduced mass between $\alpha$ particle and the daughter nucleus with $m_d$ and $m_\alpha$ being the mass of the daughter nucleus and $\alpha$ particle, respectively. $\hbar$ is the reduced Planck constant. $R_n$ = $\sqrt{\frac{3}{5}} R_{00}$ denotes the nucleus root-mean-square (rms) radius with $R_{00}$ = $1.240 A^{1/3} (1 + \frac{1.646} {A}-0.191 \frac{A - 2 Z}{A})$ \cite{PhysRevC.62.044610}, where $A$ and $Z$ are the proton and mass number of parent nucleus. $G = 2n_r + l$ represents the main quantum number with $n_r$ and $l$ being the radial quantum number and the angular quantity quantum number, respectively. For $\alpha$ decay, \emph{G} can be obtained by \cite{PhysRevC.69.024614}
\begin{equation}
\label{eq7}
\
G = 2n_r + l = \left\{\begin{array}{llll}
18, &N\leq 82, \\
20, &82<N\leq 126, \\
22, &N>126.
\end{array}\right.
\end{equation}
$l_{\rm{min}}$,\;the minimum angular momentum taken away by the $\alpha$ particle, can be obtained by \cite{PhysRevC.82.059901}
\begin{equation}
\label{eq7}
\
l_{\rm{min}} = \left\{\begin{array}{llll}
\Delta _j,            & \mbox{for even}\;{\Delta _j}\;{and}\;{\pi_p} = {\pi_d}, \\
{\Delta_ j} + 1,    & \mbox{for even}\;{\Delta _j}\;{and}\;{\pi_p} \ne {\pi_d}, \\
\Delta _j,            & \mbox{for odd}\;{\Delta _j} \;{and} \;{\pi_p} \ne {\pi_d}, \\
{\Delta_ j} + 1,    & \mbox{for odd}\; {\Delta _j}\;{and}\; {\pi_p} = {\pi_d},
\end{array}\right.
\end{equation}
where $\Delta _j = \left \vert{j_p - j_d}\right \vert$. $j_p$, $\pi_p$, $j_d$, $\pi_d$ represent spin and parity values of the parent and daughter
nuclei, respectively.\\

The generalized liquid drop model (GLDM) has been successfully employed to describe the fusion reactions \cite{ROYER1985477} and nuclear decays \cite{Bao_2012,BAO20131,PhysRevC.74.017304}. Within the GLDM, the macroscopic total energy $E$ is defined as \cite{Royer_2000}
\begin{equation}
E = E_S + E_V + E_C + E_{\rm{prox}},
\end {equation}
where $E_S$, $E_V$, $E_C$ and $E_{\rm{prox}}$ are the surface, volume, Coulomb and proximity energies, respectively. For one-body shapes, the $E_S$, $E_V$ and $E_C$ can be expressed as
\begin{eqnarray}
E_S = 17.9439 (1 - 2.6 I^2) A ^{2/3} (S/4\pi {R_0^2}) \,\rm{MeV}, \notag \\
E_V = -15.494 (1 - 1.8 I^2) A \,\rm{MeV},  \\
E_C = 0.6 e^2(Z^2/R_0) \times 0.5\int(V(\theta) /V_0 )(R(\theta)/R_0)^3 \sin\theta \,d\theta, \notag 
\end{eqnarray}
where $I = (N-Z)/A$ is the relative neutron excess. $S$ is the surface of the deformed nucleus. $V(\theta)$ is the electrostatic potential at the surface and $V_0$ is the surface potential of the sphere. ${R_0} = 1.28 A^{1/3} - 0.76+0.8A^{-1/3}$ is the effective sharp radius\cite{BLOCKI1977427}.

For two separated spherical nuclei, the $E_S$,  $E_V$ and $E_C$ are defined as:
\begin{eqnarray}
E_S = 17.9439 [(1 - 2.6{I_1}^2) {A_1}^{2/3} + (1-2.6{I_2}^2){A_2}^{2/3}]\,\rm{MeV},\notag\\
E_V = -15.494 [(1 - 1.8 {I_1}^2) A_1 + (1 - 1.8 {I_2}^2) A_2] \,\rm{MeV}, \\
E_C = 0.6 e^2{Z_1}^2/R_1 + 0.6 e^2{Z_2}^2/R_2 + e^2 Z_1 Z_2/r \,\rm{MeV},\notag  
\end{eqnarray}
where $A_i$, $Z_i$ and $I_i$ are the mass number, charge number and relative neutron excesses of these nuclei, respectively. r is the
distance between the mass centres. $R_1$ and $R_2$ are the radii of the daughter nuclei and the $\alpha$ particle, which can obtained by the following relationship

\begin{eqnarray}
R_1 =R_0 (1 + \beta^{3})^{-1/3},\\
R_2 =R_0 \beta (1 + \beta^{3})^{-1/3},\notag
\end{eqnarray}
where
\begin{equation}
\beta = \frac{1.28 {A_2}^{1/3} - 0.76 + 0.8 {A_2}^{-1/3}}{1.28 {A_1}^{1/3} - 0.76 + 0.8 {A_1}^{-1/3}}. 
\end{equation}

The surface energy $E_S$ comes from the effects of the surface tension forces in a half-space and does not include the contribution of the attractive nuclear forces between the considered surfaces in the neck or in the gap between the fragments. The nuclear proximity energy term $E_{\rm{prox}}$ has been introduced to take into account these additional surface effects. It can be defined as
\begin{equation}
E_{\rm{prox}}(r) = 2\gamma \int _{h_{\rm{min}}}^{h_{\rm{max}}} \Phi{[D(r,h)/b]}2 \pi h \, dh. 
\end {equation}
Here $h$ is the transverse distance varying from the neck radius or zero to the height of the neck border. After the separation, $h_{\rm{min}}$ = 0 and $h_{\rm{max}}$ = $R_2$. $D$ is the distance between the opposite infinitesimal surfaces considered, $b$ is the surface width fixed at the standard value of 0.99 fm, $\Phi$ is the proximity function of Feldmeier, $\gamma = 0.9517\sqrt{(1 - 2.6 {I_1}^2) (1 - 2.6 {I_2}^2)}$ MeV ${\rm{fm}}^{-2}$ is the geometric mean between the surface parameters of the two nuclei. 

\emph{P}, the penetration probability of the $\alpha$ particle crossing the barrier, is calculated within the action integral
\begin{equation}
P = \rm{exp} \left[ -\frac {2} {\hbar} \int _{\emph{R}_{\rm{in}}}^{\emph{R}_{\rm{out}}} \sqrt{2 \emph{B}(\emph{r}) (\emph{E}(\emph{r})- \emph{E}(sphere))}\,dr\right].
\label{subeq:11}
\end {equation}
Here, $B(r) = \mu$, $Q_{\alpha}$ is the $\alpha$ decay energy. $R_{\rm{in}}$ and $R_{\rm{out}}$ are the two turning points of the semiclassical Wentzel-Kramers-Brillouin (WKB) action integral with $E(R_{\rm{in}}) = E(R_{\rm{out}} ) = Q_{\alpha}$. $R_{\rm{in}} = R_1 + R_2$. Consisdering the contribution of the centrifugal potential, $R_{\rm{out}}$ can be obtained by
\begin{equation}
R_{\rm{out}} = \frac{Z_1 Z_2 e^2}{2 Q_{\alpha}} + \sqrt{{\left(\frac{Z_1 Z_2 e^2} {2 Q_{\alpha}}\right)}^2 + \frac{{l (l + 1) {\hbar}^2}}{2 \mu Q_{\alpha}}}.
\end{equation}
\section{Results and discussion}

In recent years, many works have been indicated that the $\alpha$ preformation factor $P_\alpha$ of the nucleus becomes smaller as the valence nucleus (hole) becomes smaller \cite{PhysRevC.84.027303,Qian2013,PhysRevC.80.064325}. Meanwhile, intensive works related to the $N_pN_n$ scheme have indicated the importance of the proton-neutron interaction in determining the evolution of nuclear structure \cite{PhysRev.92.1211,RevModPhys.34.704,FEDERMAN19799,PhysRevLett.54.1991,PhysRevLett.58.658}. In 2011, Seif \emph{et \ al.} found that the $P_\alpha$ are linearly related to the multiplication of valence proton numbers and valence neutron numbers $N_p$$N_n$ for even-even nuclei around $\emph{Z}$ = 82, $\emph{N}$ = 126 shell closures \cite{PhysRevC.84.064608}. Very recently, we found that this linear relationship is still exist in the cases of odd-$\emph{A}$ and doubly-odd nuclei for favored and unfavored $\alpha$ decay\cite{PhysRevC.94.024338,PhysRevC.96.024318,PhysRevC.97.044322}. Moreover, we systematically studied the $P_\alpha$ based on the cluster-formation model (CFM)\cite{Saleh_Ahmed_2013,Deng_2015,PhysRevC.93.044326,SALEHAHMED2017103} of all kinds of nuclei around $\emph{Z}$ = 82, $\emph{N}$ = 126 shell closures. The results indicated that the linear relationship between the $\alpha$ preformation factors and $N_pN_n$ still exists. Whether this phenomenon is model dependent or may the effect of the valance proton-neutron interaction around the closed shells is an interesting problem. In present work,\;we use the generalized liquid drop model (GLDM) to systematically study the $P_\alpha$ and $\alpha$ decay half-lives of nuclei around $\emph{Z}$ = 82, $\emph{N}$ = 126 shell closures, while the experimental $\alpha$ decay half-lives $T^{\rm{expt}}_{1/2}$ are taken from the latest evaluated nuclear properties table NUBASE2016 \cite{Audi_2017}.

First, we calculate the $\alpha$ preformation factors $P_\alpha$ of 152 nuclei (including 47 even-even nuclei, 73 odd-$\emph{A}$ nuclei, and 32 doubly-odd nuclei) around $\emph{Z}$ = 82, $\emph{N}$ = 126 shell closures through $P_{\alpha} = \frac{T^{\rm{calc}}_{1/2}}{T^{\rm{expt}}_{1/2}}$, while the theoretical calculations ${T^{\rm{calc}}_{1/2}}$ are obtained by GLDM. The calculated results of $P_\alpha$ are listed the fifth column of Tables \ref{table 1}-\ref{table 5}. For more clearly observe the relationship between the obtained $P_\alpha$ and $N_pN_n$ for these nuclei, we plot the $P_\alpha$ as the function of $\frac{N_pN_n}{Z_0 + N_0}$ in five cases in Figs.\,\ref{fig 1}-\ref{fig 3} (Fig.\,\ref{fig 1} for the case of even-even nuclei, Fig.\,\ref{fig 2} for the cases of the favored and unfavored odd-$\emph{A}$ nuclei, Fig.\,\ref{fig 3} for the cases of the favored and unfavored doubly-odd nuclei), respectively. To more clearly describe the $N_p$ and $N_n$, bounded by the double magic numbers $Z$ = 82, $N$ = 126, we divide these 152 nuclei into three regions. In region \uppercase\expandafter{\romannumeral1}, the proton number of nucleus is above the $Z$ = 82 shell closure and the neutron number is below the $N$ = 126 closed shell, while the $\frac{N_pN_n}{Z_0+N_0}$ value is negative. In region \uppercase\expandafter{\romannumeral2}, the proton number of nucleus is above the $Z$ = 82 shell closure and the neutron number is above the $N$ = 126 closed shell, while the $\frac{N_pN_n}{Z_0 + N_0}$ value is positive. In region \uppercase\expandafter{\romannumeral3}, the proton number of nucleus is below the $Z$ = 82 shell closure and the neutron number is above the $N$ = 126 closed shell, while the $\frac{N_pN_n}{Z_0 + N_0}$ value is positive. In Fig.\,\ref{fig 1}, the red dots represents the case of even-even nuclei $\alpha$ decay, the red dash line represents the fittings of $P_\alpha$ for cases of even-even nuclei $\alpha$ decay. In Figs.\,\ref{fig 2}-\ref{fig 3}, the red dots and blue triangle represent the cases of favored and unfavored $\alpha$ decay, respectively.\;The red dash and blue solid lines represent the fittings of $P_\alpha$. As we can see in Figs.\,\ref{fig 1}-\ref{fig 3}, around $\emph{Z}$ = 82, $\emph{N}$ = 126 shell closures, when the values of $\frac{N_pN_n}{Z_0 + N_0}$ are positive, $P_\alpha$ increases basically with the increase of $\frac{N_pN_n}{Z_0 + N_0}$. Similarly, when the values of $\frac{N_pN_n}{Z_0 + N_0}$ are negative, $P_\alpha$ increases basically with the increase of $\frac{N_pN_n}{Z_0 + N_0}$. Furthermore, we find that the linear relationships of Figs.\,\ref{fig 1}-\ref{fig 3} are relatively obvious except the right sides of Figs.\,\ref{fig 2}-\ref{fig 3} for the cases of favored decay. Obviously, some red dots in the right sides of Figs.\,\ref{fig 2}-\ref{fig 3} deviate too much from the fitted line. Careful analysis of the right side of Fig.\,\ref{fig 2}, a fantastic phenomenon appears: the $P_\alpha$ values of some odd-$Z$ nuclei are even larger than ones of most even-even nuclei i.e. $P_\alpha$ = 0.572 for $^{217}$Fr, 0.534 for $^{217}$Ac, 0.826 for $^{219}$Pa. Theoretically, for odd-$A$ and odd-odd nucleus, due to the Pauli-blocking effect, the unpaired nucleon(s) would lead to the $\alpha$ particle formation suppressed, thus the $P_\alpha$ of odd-$A$ should not be too large. In response to aboved phenomenon of the right side of Fig.\,\ref{fig 2}, we note that the neutron number $N$ of $^{217}$Fr, $^{217}$Ac and $^{219}$Pa are 130, 128 and 128, respectively, which are near the magic number $N$ = 126. Recent works showed that the $P_\alpha$ values decrease with increasing neutron number until the shell closure at $N$ = 126, then sharp increase with $N$, implying the important influence of shell effects on $\alpha$ particle preformation process in the parent nuclei \cite{Zhang2009, Zhao2018}. As we can see from the right side of Fig.\,\ref{fig 3}, there are seven red dots for favored odd-odd nuclei decay, however, we can find five of them have uncertain spin or parity from table\ref{table 4} i.e. $^{216}$At, $^{216}$Fr. Hence, the inaccuracy of minimum angular momentum taken away by the $\alpha$ particle ($l_{\rm{min}}$) will in turn affects the $P_\alpha$ values extracted from the radio of calculated decay half-life to experimental ones.

In 2018, based on the single particle energy spectra obtained by the relativistic Hartree-Bogoliubov mean field model \cite{PhysRevC.76.034314}, Sun and Zhang defined the microscopic valence nucleon (holes) numbers (${\Omega}_{\pi}, {\Omega}_{\nu}$) and proposed two different formulas to calculate $P_\alpha$, which can be expressed as \cite{Sun_2018}

\begin{equation}
P_\alpha = a\,{\Omega}_{\pi} {\Omega}_{\nu}\,\{1 + b\,\rm{exp} [-\frac{({\lambda}_{\pi} - {\lambda}_{\nu})^2} {2 {\sigma}^2}]\},
\label{subeq:14}
\end {equation}

\begin{equation}
P_\alpha = a_{\rm{nn},\rm{nh}}\,{\Omega}_{\pi} {\Omega}_{\nu}.
\label{subeq:15}
\end {equation}
Here parameters $a$, $b$ and $\sigma$ in eq.\,(\ref{subeq:14}), as well as $a_{\rm{nn},\rm{nh}}$ in eq.\,(\ref{subeq:15}) correspond to the valence proton-neutron interaction strength.

In their work, using the aboved formula, they systematically investigated the $\alpha$ decay preformation factors $P_\alpha$ for even-even polonium, radon, radium and thorium isotopes. Based on the calculated results of their work, we plot $P_\alpha$ as the function of $\frac{N_pN_n}{Z_0 + N_0}$ in Figs.\,\ref{fig 4}-\ref{fig 5}. In these figures, the blue opened squares represent the even-even nuclei extracted from Sun and Zhang work, which are located around shell closures, the blue dashed lines represent the fittings of the $P_{\alpha}$. As we can see from these figures, like Fig.\,\ref{fig 1}, for the case of even-even nuclei, the $P_{\alpha}$ obtained by eq.\,(\ref{subeq:14}) and eq.\,(\ref{subeq:15}) also have an obvious linear relationship with $\frac{N_pN_n}{Z_0 + N_0}$. This phenomenon further implies that this linear relationship is not model dependent. Furthermore, for the cases of unfavored $\alpha$ decay, from Figs.\,\ref{fig 2}-\ref{fig 3}, we find that the values of $P_\alpha$ are relatively small relative to the favored $\alpha$ decay. The reasons may be the influence of centrifugal potential, which reduces the $\alpha$ decay width or nuclear structure configuration changes. Combining with the analysis results of Figs.\,\ref{fig 1}-\ref{fig 3}, we suspect the linear relationship between $P_\alpha$ and $N_pN_n$ for all kinds of nuclei may be related to the effect of valence proton-neutron around shell closures. In order to deeper study the relationship between $\alpha$ preformation factors $P_\alpha$ and $N_pN_n$, all cases of $P_\alpha$ can be obtained by the linear relationship,
\begin{equation}
P_\alpha = a \frac{N_pN_n}{Z_0 + N_0}+b.
\label{subeq:16}
\end{equation}
Here $a$ and $b$ are adjustable parameters which are extracted from fittings of Figs.\,\ref{fig 1}-\ref{fig 3} and listed in Table\,\ref{table 6}. In theory, combined on these parameters and Eq.\,(\ref{subeq:15}), one can obtain the fitted $P_\alpha$.

\begin{figure}[htbp]
\centering
\subfigure[]    {
\includegraphics[width=8cm]{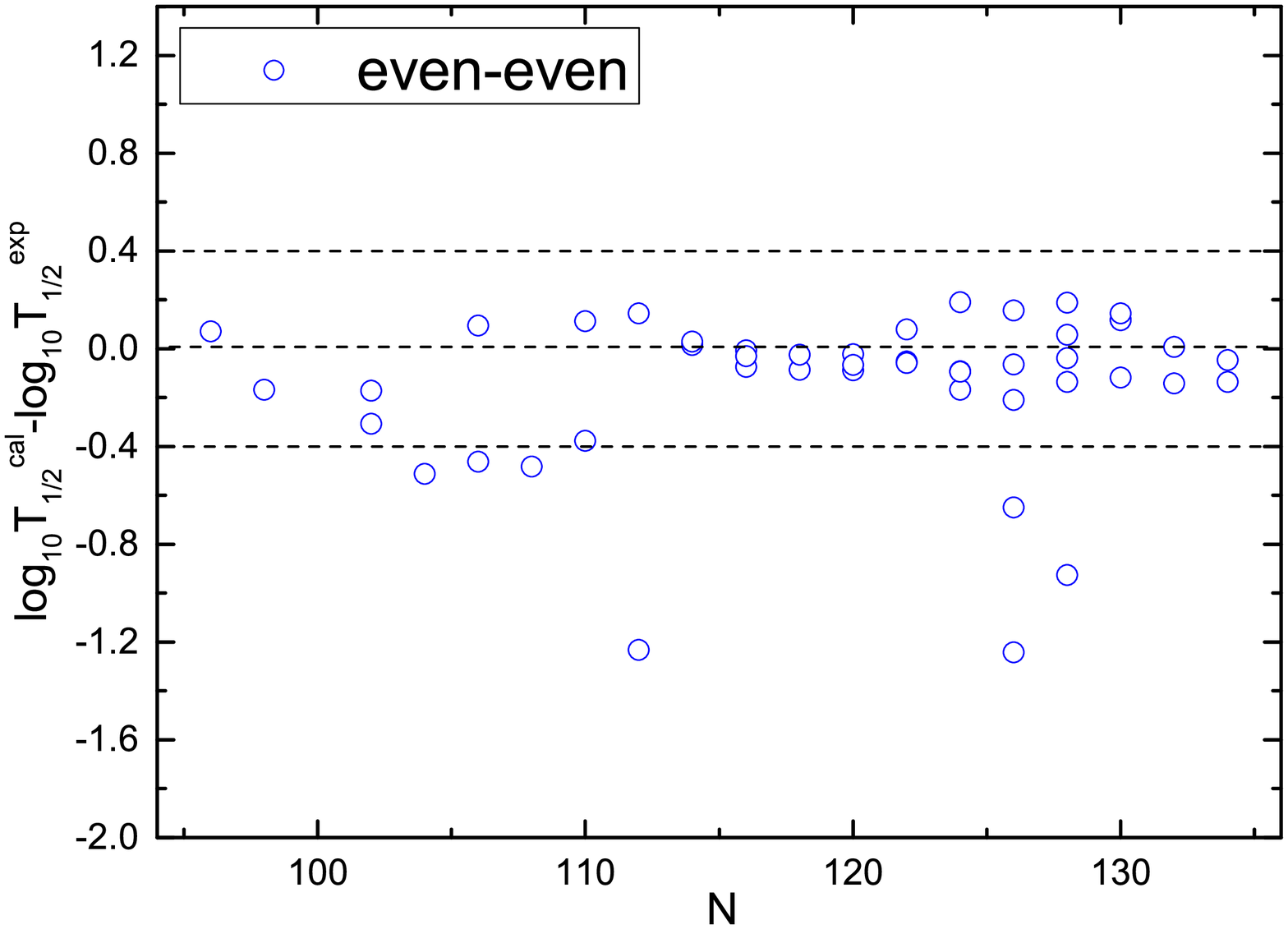}
\label {fig 6(a)}
}
\quad
\subfigure[]{
\includegraphics[width=8cm]{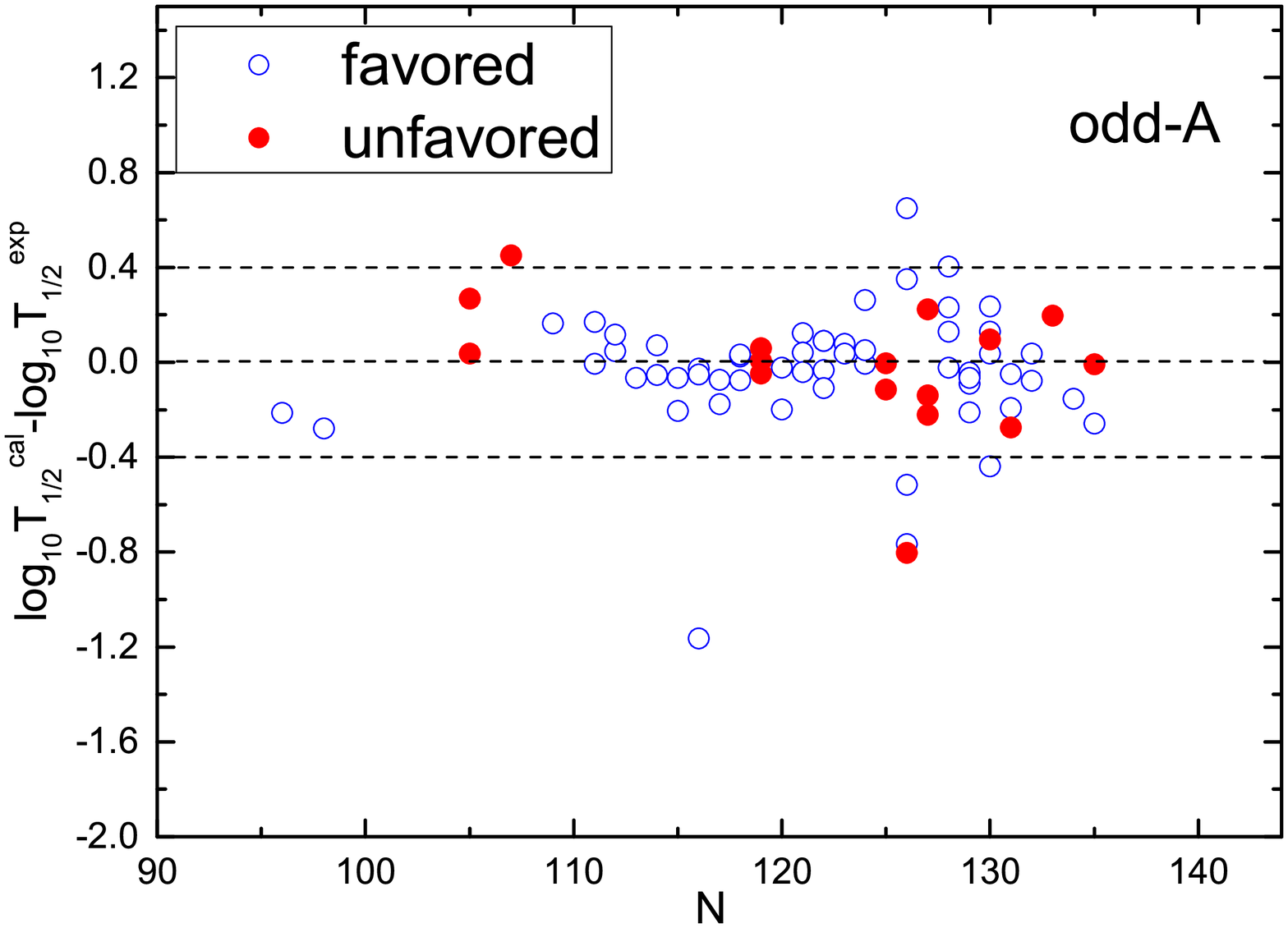}
\label {fig 6(b)}
}
\quad
\subfigure[]{
\includegraphics[width=8cm]{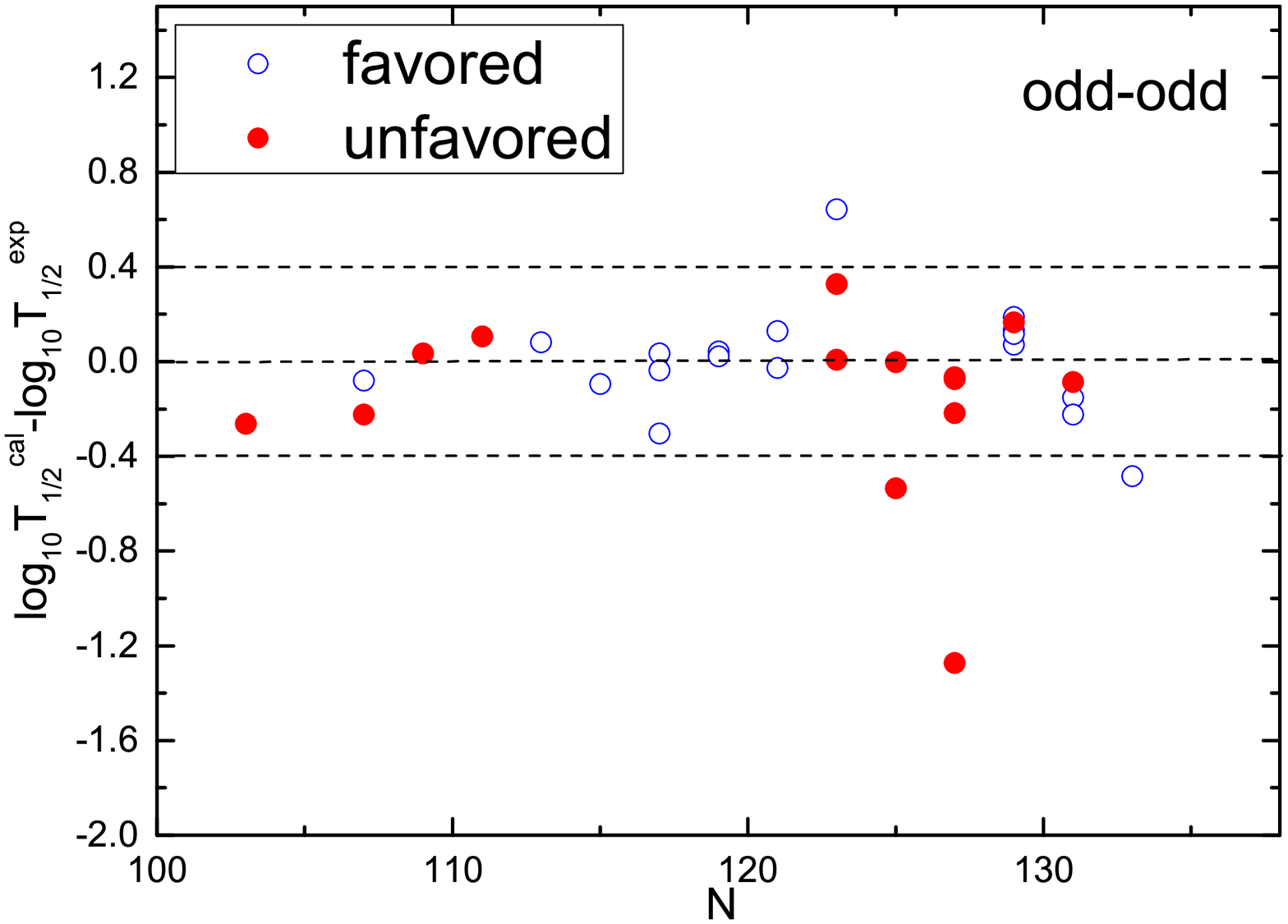}
\label {fig 6(c)}
}
\caption{The deviations of logarithmic form of $\alpha$ decay half-lives between calculations using GLDM with fitting $P_\alpha$ calculated by Eq.\,(\ref{subeq:16}) and experimental data as a function of neutron numbers.}
\label {fig 6}
\end{figure}

Second, we systematically calculate the $\alpha$ decay half-lives of 152 nuclei using GLDM. All the numerical results are shown in Tables \ref{table 1}-\ref{table 5}. In these tables, the first five columns denote the $\alpha$ transition, $\alpha$ decay energy $Q_\alpha$ which is taken from the latest evaluated atomic mass table AME2016 \cite{Wang_2017,Huang_2017}, spin-parity transformation (${j^{\pi}_{p}}\to{j^{\pi}_{d}}$), minimum orbital angular momentum $l_{\rm{min}}$ taken away by the $\alpha$ particle and the extracted $\alpha$ preformation factor, respectively. The sixth column denotes the logarithmic form of the experimental $\alpha$ decay half-life denoted as ${\rm{lg}T_{1/2}^{\rm{exp}}}$. The last two columns denote  the logarithmic form of calculated $\alpha$ decay half-life using GLDM without considering $P_\alpha$ and with fitting $P_\alpha$ calculated by Eq.\,(\ref{subeq:15}), which are denoted as ${\rm{lg}T_{1/2}^{\rm{calc1}}}$ and ${\rm{lg}T_{1/2}^{\rm{calc2}}}$, respectively. Simultaneously, each table is divided into two parts: region \uppercase\expandafter{\romannumeral1}, regions \uppercase\expandafter{\romannumeral2} and \uppercase\expandafter{\romannumeral3}. As can be seen from Tables \ref{table 1}-\ref{table 5}, relative to ${\rm{lg}T_{1/2}^{\rm{calc1}}}$, ${\rm{lg}T_{1/2}^{\rm{calc2}}}$ can better reproduce experimental data. For more intuitively, we calculate the standard deviations $\sigma$ = $\sqrt{\sum{(\rm{log}_{10}{T_{1/2}^{\rm{calc}}} - \rm{log}_{10}{T_{1/2}^{\rm{exp}}})^2}/n}$ between the calculated $\alpha$ decay half-lives and experimental data. The calculated results are given in Table\,\ref{table 7}. In this table, $\sigma_1$, $\sigma_2$ represent the standard deviations between $T^{\rm{calc1}}_{1/2}$, $T^{\rm{calc2}}_{1/2}$ and $T^{\rm{exp}}_{1/2}$, respectively. As we can see from Table\,\ref{table 7}, it is obviously that for all kinds of nuclei, the values of $\sigma_1$ is much bigger than the values of $\sigma_2$. The maximum value of $\sigma_2$ is only 0.40, it implies that the calculated $\alpha$ decay half-lives using the GLDM with fitting $P_\alpha$ calculated by Eq.\,(\ref{subeq:15}) can better reproduce the experimental data. Moreover, for more clearly indicating the agreement between the calculations using the GLDM with the fitting $P_\alpha$ and the experimental data, we plot the logarithm deviation between the calculated results and experimental data for all kinds of nuclei in Figs.\,\ref{fig 6(a)}-\ref{fig 6(c)}. In Figs.\,\ref{fig 6(a)}-\ref{fig 6(c)}, the opened blue circles denote the logarithm deviation between the calculated results and experimental data of the cases of favored $\alpha$ decay for all kinds of nuclei. The red circles in Figs.\,\ref{fig 6(b)}-\ref{fig 6(c)} denote the cases of unfavored $\alpha$ decay for odd-A and odd-odd nuclei, respectively. One can see from Figs.\,\ref{fig 6(a)}-\ref{fig 6(c)}, the values of $\rm{log}_{10}T_{1/2}^{\rm{cal}} - \rm{log}_{10}T_{1/2}^{\rm{exp}}$ are basically between -0.4 and 0.4. It indicates that GLDM with the fitting $P_\alpha$ can be treated as a useful tool to study the $\alpha$ decay half-lives of nuclei around the $\emph{Z}$ = 82, $\emph{N}$ = 126 shell closures. 
\end{multicols}

\begin{center}
\tabcaption{Calculations of $\alpha$ decay half-lives in logarithmic form and the $\alpha$ preformation factors of even-even nuclei in Region I-III around $Z = 82$, $N = 126$ closed shells. The experimental $\alpha$ decay half-lives, spin and parity are taken from the latest evaluated nuclear properties table NUBASE2016 \cite{Audi_2017}, the $\alpha$ decay energies ${Q_{\alpha}}$ are taken from the latest evaluated atomic mass table AME2016 \cite{Wang_2017,Huang_2017}. The $\alpha$ preformation factors ${P_{\alpha}}$ are extracted from the ratios of calculated $\alpha$ decay half-lives to the experimental data\cite{PhysRevC.84.027303,PhysRevC.80.064325,Qian2013}, while the calculated $\alpha$ decay half-lives are obtained by GLDM.}
\label {table 1}
\footnotesize
\begin{tabular}{cccccccc}
\hline {$\mathcal{\alpha}$ transition} & $Q_{\alpha}$ (MeV) & ${j^{\pi}_{p}}\to{j^{\pi}_{d}}$ &$l_{\rm{min}}$ &${P_{\alpha}}$ & $\rm{lg}T^{\rm{expt}}_{1/2}$ (s)&${\rm{lg}T_{1/2}^{\rm{calc1}}}$ (s)& ${\rm{lg}T_{1/2}^{\rm{calc2}}}$ (s)\\ \hline
 \noalign{\global\arrayrulewidth1pt}\noalign{\global\arrayrulewidth0.4pt} \multicolumn{8}{c}{\textbf{Nuclei in Region I}}\\
$	^{	186	}	$	Po	$	\to	^{	182	}$	Pb	$$&	8.501 	&${0^+}\to{0^+}$ &0&	0.126 	&$	-4.469 	$&$	-5.369 	$&$	-4.641 	$\\
$	^{	190	}	$	Po	$	\to	^{	186	}$	Pb	$$&	7.693 	&${0^+}\to{0^+}$ &0&	0.201 	&$	-2.609 	$&$	-3.307 	$&$	-2.515 	$\\
$	^{	194	}	$	Po	$	\to	^{	190	}$	Pb	$$&	6.987 	&${0^+}\to{0^+}$ &0&	0.176 	&$	-0.407 	$&$	-1.160 	$&$	-0.293 	$\\
$	^{	196	}	$	Po	$	\to	^{	192	}$	Pb	$$&	6.658 	&${0^+}\to{0^+}$ &0&	0.172 	&$	0.754 	$&$	-0.011 	$&$	0.899 	$\\
$	^{	198	}	$	Po	$	\to	^{	194	}$	Pb	$$&	6.310 	&${0^+}\to{0^+}$ &0&	0.114 	&$	2.266 	$&$	1.324 	$&$	2.282 	$\\
$	^{	200	}	$	Po	$	\to	^{	196	}$	Pb	$$&	5.981 	&${0^+}\to{0^+}$ &0&	0.082 	&$	3.793 	$&$	2.707 	$&$	3.719 	$\\
$	^{	202	}	$	Po	$	\to	^{	198	}$	Pb	$$&	5.700 	&${0^+}\to{0^+}$ &0&	0.069 	&$	5.143 	$&$	3.984 	$&$	5.057 	$\\
$	^{	204	}	$	Po	$	\to	^{	200	}$	Pb	$$&	5.485 	&${0^+}\to{0^+}$ &0&	0.059 	&$	6.275 	$&$	5.043 	$&$	6.187 	$\\
$	^{	206	}	$	Po	$	\to	^{	202	}$	Pb	$$&	5.327 	&${0^+}\to{0^+}$ &0&	0.052 	&$	7.144 	$&$	5.862 	$&$	7.091 	$\\
$	^{	208	}	$	Po	$	\to	^{	204	}$	Pb	$$&	5.216 	&${0^+}\to{0^+}$ &0&	0.031 	&$	7.961 	$&$	6.457 	$&$	7.793 	$\\
$	^{	200	}	$	Rn	$	\to	^{	196	}$	Po	$$&	7.043 	&${0^+}\to{0^+}$ &0&	0.200 	&$	0.070 	$&$	-0.629 	$&$	0.099 	$\\
$	^{	202	}	$	Rn	$	\to	^{	198	}$	Po	$$&	6.773 	&${0^+}\to{0^+}$ &0&	0.159 	&$	1.090 	$&$	0.291 	$&$	1.083 	$\\
$	^{	204	}	$	Rn	$	\to	^{	200	}$	Po	$$&	6.547 	&${0^+}\to{0^+}$ &0&	0.128 	&$	2.012 	$&$	1.120 	$&$	1.987 	$\\
$	^{	206	}	$	Rn	$	\to	^{	202	}$	Po	$$&	6.384 	&${0^+}\to{0^+}$ &0&	0.105 	&$	2.737 	$&$	1.757 	$&$	2.715 	$\\
$	^{	208	}	$	Rn	$	\to	^{	204	}$	Po	$$&	6.260 	&${0^+}\to{0^+}$ &0&	0.074 	&$	3.367 	$&$	2.236 	$&$	3.309 	$\\
$	^{	210	}	$	Rn	$	\to	^{	206	}$	Po	$$&	6.159 	&${0^+}\to{0^+}$ &0&	0.048 	&$	3.954 	$&$	2.631 	$&$	3.861 	$\\
$	^{	212	}	$	Rn	$	\to	^{	208	}$	Po	$$&	6.385 	&${0^+}\to{0^+}$ &0&	0.029 	&$	3.157 	$&$	1.616 	$&$	3.093 	$\\
$	^{	204	}	$	Ra	$	\to	^{	200	}$	Rn	$$&	7.637 	&${0^+}\to{0^+}$ &0&	0.210 	&$	-1.222 	$&$	-1.900 	$&$	-1.253 	$\\
$	^{	208	}	$	Ra	$	\to	^{	204	}$	Rn	$$&	7.273 	&${0^+}\to{0^+}$ &0&	0.128 	&$	0.104 	$&$	-0.789 	$&$	0.039 	$\\
$	^{	214	}	$	Ra	$	\to	^{	210	}$	Rn	$$&	7.273 	&${0^+}\to{0^+}$ &0&	0.048 	&$	0.387 	$&$	-0.933 	$&$	0.543 	$\\
$	^{	212	}	$	Th	$	\to	^{	208	}$	Ra	$$&	7.958 	&${0^+}\to{0^+}$ &0&	0.163 	&$	-1.499 	$&$	-2.287 	$&$	-1.420 	$\\
$	^{	214	}	$	Th	$	\to	^{	210	}$	Ra	$$&	7.827 	&${0^+}\to{0^+}$ &0&	0.131 	&$	-1.060 	$&$	-1.942 	$&$	-0.869 	$\\
$	^{	216	}	$	U	$	\to	^{	212	}$	Th	$$&	8.530 	&${0^+}\to{0^+}$ &0&	0.078 	&$	-2.161 	$&$	-3.266 	$&$	-2.255 	$\\
		 \noalign{\global\arrayrulewidth1pt}\noalign{\global\arrayrulewidth0.4pt} \multicolumn{8}{c}{\textbf{Nuclei in Regions II and III}}\\								
$	^{	178	}	$	Pb	$	\to	^{	174	}$	Hg	$$&	7.790 	&${0^+}\to{0^+}$ &0&	0.380 	&$	-3.638 	$&$	-4.059 	$&$	-3.568 	$\\
$	^{	180	}	$	Pb	$	\to	^{	176	}$	Hg	$$&	7.419 	&${0^+}\to{0^+}$ &0&	0.220 	&$	-2.387 	$&$	-3.045 	$&$	-2.554 	$\\
$	^{	184	}	$	Pb	$	\to	^{	180	}$	Hg	$$&	6.773 	&${0^+}\to{0^+}$ &0&	0.160 	&$	-0.213 	$&$	-1.011 	$&$	-0.520 	$\\
$	^{	186	}	$	Pb	$	\to	^{	182	}$	Hg	$$&	6.470 	&${0^+}\to{0^+}$ &0&	0.099 	&$	1.072 	$&$	0.069 	$&$	0.560 	$\\
$	^{	188	}	$	Pb	$	\to	^{	184	}$	Hg	$$&	6.109 	&${0^+}\to{0^+}$ &0&	0.111 	&$	2.427 	$&$	1.474 	$&$	1.965 	$\\
$	^{	190	}	$	Pb	$	\to	^{	186	}$	Hg	$$&	5.697 	&${0^+}\to{0^+}$ &0&	0.107 	&$	4.245 	$&$	3.273 	$&$	3.763 	$\\
$	^{	192	}	$	Pb	$	\to	^{	188	}$	Hg	$$&	5.221 	&${0^+}\to{0^+}$ &0&	0.136 	&$	6.546 	$&$	5.680 	$&$	6.170 	$\\
$	^{	194	}	$	Pb	$	\to	^{	190	}$	Hg	$$&	4.738 	&${0^+}\to{0^+}$ &0&	0.019 	&$	10.234 $&$	8.511   $&$	9.002 	$\\
$	^{	210	}	$	Pb	$	\to	^{	206	}$	Hg	$$&	3.793 	&${0^+}\to{0^+}$ &0&	0.038 	&$	16.967 $&$	15.550 $&$16.040  $\\
$	^{	210	}	$	Po	$	\to	^{	206	}$	Pb	$$&	5.408 	&${0^+}\to{0^+}$ &0&	0.018 	&$	7.078 	$&$	5.344 	$&$	5.835 	$\\
$	^{	212	}	$	Po	$	\to	^{	208	}$	Pb	$$&	8.954 	&${0^+}\to{0^+}$ &0&	0.247 	&$	-6.531 	$&$	-7.138 	$&$	-6.666 	$\\
$	^{	214	}	$	Po	$	\to	^{	210	}$	Pb	$$&	7.834 	&${0^+}\to{0^+}$ &0&	0.268 	&$	-3.786 	$&$	-4.358 	$&$	-3.903 	$\\
$	^{	216	}	$	Po	$	\to	^{	212	}$	Pb	$$&	6.907 	&${0^+}\to{0^+}$ &0&	0.262 	&$	-0.839 	$&$	-1.420 	$&$	-0.981 	$\\
$	^{	218	}	$	Po	$	\to	^{	214	}$	Pb	$$&	6.115 	&${0^+}\to{0^+}$ &0&	0.276 	&$	2.269 	$&$	1.711 	$&$	2.134 	$\\
$	^{	214	}	$	Rn	$	\to	^{	210	}$	Po	$$&	9.208 	&${0^+}\to{0^+}$ &0&	0.321 	&$	-6.569 	$&$	-7.062 	$&$	-6.606 	$\\
$	^{	216	}	$	Rn	$	\to	^{	212	}$	Po	$$&	8.198 	&${0^+}\to{0^+}$ &0&	0.497 	&$	-4.347 	$&$	-4.650 	$&$	-4.227 	$\\
$	^{	218	}	$	Rn	$	\to	^{	214	}$	Po	$$&	7.263 	&${0^+}\to{0^+}$ &0&	0.412 	&$	-1.472 	$&$	-1.857 	$&$	-1.464 	$\\
$	^{	220	}	$	Rn	$	\to	^{	216	}$	Po	$$&	6.405 	&${0^+}\to{0^+}$ &0&	0.388 	&$	1.745 	$&$	1.334 	$&$	1.698 	$\\
$	^{	216	}	$	Ra	$	\to	^{	212	}$	Rn	$$&	9.526 	&${0^+}\to{0^+}$ &0&	0.416 	&$	-6.740 	$&$	-7.121 	$&$	-6.682 	$\\
$	^{	218	}	$	Ra	$	\to	^{	214	}$	Rn	$$&	8.546 	&${0^+}\to{0^+}$ &0&	0.530 	&$	-4.599 	$&$	-4.874 	$&$	-4.481 	$\\
$	^{	216	}	$	Th	$	\to	^{	212	}$	Ra	$$&	8.072 	&${0^+}\to{0^+}$ &0&	0.073 	&$	-1.585 	$&$	-2.724 	$&$	-2.234 	$\\
$	^{	218	}	$	Th	$	\to	^{	214	}$	Ra	$$&	9.849 	&${0^+}\to{0^+}$ &0&	0.582 	&$	-6.932 	$&$	-7.167 	$&$	-6.744 	$\\
$	^{	220	}	$	Th	$	\to	^{	216	}$	Ra	$$&	8.953 	&${0^+}\to{0^+}$ &0&	0.604 	&$	-5.013 	$&$	-5.232 	$&$	-4.868 	$\\
$	^{	218	}	$	U	$	\to	^{	214	}$	Th	    $$&	8.775 	&${0^+}\to{0^+}$ &0&	0.199 	    &$	-3.260 	$&$	-3.961 	$&$	-3.470 	$\\ 
\hline
\end{tabular}
\end{center}

\begin{center}
\tabcaption{Same as Table \ref{table 1}, but for favored $\alpha$ decay of odd-$\emph{A}$ nuclei around the $\emph{Z}$ = 82, $\emph{N}$ = 126 shell closures. ``()'' represents uncertain spin and/or parity, and ``\#'' represents values estimated from trends in neighboring nuclides with the same $\emph{Z}$ and $\emph{N}$ parities.}
\label {table 2}
\footnotesize
\begin{tabular}{cccccccc}
\hline {$\mathcal{\alpha}$ transition} & $Q_{\alpha}$ (MeV) & ${j^{\pi}_{p}}\to{j^{\pi}_{d}}$ &$l_{\rm{min}}$ &${P_{\alpha}}$ & $\rm{lg}T^{\rm{expt}}_{1/2}$ (s)&${\rm{lg}T_{1/2}^{\rm{calc1}}}$ (s)& ${\rm{lg}T_{1/2}^{\rm{calc2}}}$ (s)\\ \hline
\noalign{\global\arrayrulewidth1pt}\noalign{\global\arrayrulewidth0.4pt} \multicolumn{8}{c}{\textbf{Nuclei in Region I}}\\
$	^{	195	}	$Po		$	\to	^{	191	}$	Pb	$$&	6.745 	&${(3/2^-)}\to{(3/2^-)}$&0&	0.100 	&$	0.692 	$&$	-0.309 	$&$	0.687 	$\\
$	^{	197	}	$Po		$	\to	^{	193	}$	Pb	$$&	6.405 	&${(3/2^-)}\to{(3/2^-)}$&0&	0.076 	&$	2.079 	$&$	0.960 	$&$	2.014 	$\\
$	^{	199	}	$Po		$	\to	^{	195	}$	Pb	$$&	6.075 	&${(3/2^-)}\to{3/2^-}$&0&	0.047 	&$	3.639 	$&$	2.314 	$&$	3.434 	$\\
$	^{	201	}	$Po		$	\to	^{	197	}$	Pb	$$&	5.799 	&${3/2^-}\to{3/2^-}$&0&	0.042 	&$	4.917 	$&$	3.541 	$&$	4.740 	$\\
$	^{	205	}	$Po		$	\to	^{	201	}$	Pb	$$&	5.325 	&${5/2^-}\to{5/2^-}$&0&	0.050 	&$	7.185 	$&$	5.887 	$&$	7.309 	$\\
$	^{	207	}	$Po		$	\to	^{	203	}$	Pb	$$&	5.216 	&${5/2^-}\to{5/2^-}$&0&	0.030 	&$	7.993 	$&$	6.473 	$&$	8.071 	$\\
$	^{	197	}	$At		$	\to	^{	193	}$	Bi$$&	7.105 	&${(9/2^-)}\to{(9/2^-)}$&0&	0.155 	&$	-0.394 	$&$	-1.205 	$&$	-0.347 	$\\
$	^{	199	}	$	At	$	\to	^{	195	}$	Bi$$&	6.778 	&${9/2^(-)}\to{9/2^(-)}$&0&	0.106 	&$	0.894 	$&$	-0.080 	$&$	0.841 	$\\
$	^{	201	}	$At		$	\to	^{	197	}$	Bi$$&	6.473 	&${(9/2^-)}\to{(9/2^-)}$&0&	0.095 	&$	2.076 	$&$	1.053 	$&$	2.048 	$\\
$	^{	203	}	$	At	$	\to	^{	199	}$	Bi$$&	6.210 	&${9/2^-}\to{9/2^-}$&0&	0.087 	&$	3.152 	$&$	2.091 	$&$	3.176 	$\\
$	^{	205	}	$	At	$	\to	^{	201	}$	Bi	$$&	6.019 	&${9/2^-}\to{9/2^-}$&0&	0.040 	&$	4.299 	$&$	2.901 	$&$	4.101 	$\\
$	^{	207	}	$At		$	\to	^{	203	}$	Bi$$&	5.873 	&${9/2^-}\to{9/2^-}$&0&	0.054 	&$	4.814 	$&$	3.549 	$&$	4.903 	$\\
$	^{	209	}	$	At	$	\to	^{	205	}$	Bi$$&	5.757 	&${9/2^-}\to{9/2^-}$&0&	0.025 	&$	5.672 	$&$	4.068 	$&$	5.666 	$\\
$	^{	211	}	$	At	$	\to	^{	207	}$	Bi$$&	5.983 	&${9/2^-}\to{9/2^-}$&0&	0.014 	&$	4.793 	$&$	2.945 	$&$	5.144 	$\\
$	^{	195	}	$	Rn	$	\to	^{	191	}$	Po	$$&	7.694 	&${3/2^-}\to{(3/2^-)}$&0&	0.322 	&$	-2.155 	$&$	-2.647 	$&$	-1.991 	$\\
$	^{	197	}	$	Rn	$	\to	^{	193	}$	Po	$$&	7.410 	&${(3/2^-)}\to{(3/2^-)}$&0&	0.289 	&$	-1.268 	$&$	-1.807 	$&$	-1.099 	$\\
$	^{	203	}	$	Rn	$	\to	^{	199	}$Po	$$&	6.629 	&${3/2^-\#}\to{(3/2^-)}$&0&	0.101 	&$	1.818 	$&$	0.825 	$&$	1.745 	$\\
$	^{	207	}	$	Rn	$	\to	^{	203	}$	Po	$$&	6.251 	&${5/2^-}\to{5/2^-}$&0&	0.076 	&$	3.417 	$&$	2.299 	$&$	3.458 	$\\
$	^{	209	}	$	Rn	$	\to	^{	205	}$	Po	$$&	6.155 	&${5/2^-}\to{5/2^-}$&0&	0.048 	&$	4.000 	$&$	2.683 	$&$	4.037 	$\\
$	^{	199	}	$	Fr	$	\to	^{	195	}$	At	$$&	7.816 	&${1/2^+\#}\to{1/2^+}$&0&	0.298 	&$	-2.180 	$&$	-2.706 	$&$	-2.063 	$\\
$	^{	201	}	$	Fr	$	\to	^{	197	}$	At	$$&	7.519 	&${(9/2^-)}\to{(9/2^-)}$&0&	0.231 	&$	-1.202 	$&$	-1.839 	$&$	-1.130 	$\\
$	^{	203	}	$	Fr	$	\to	^{	199	}$	At	$$&	7.274 	&${9/2^-}\to{9/2^(-)}$&0&	0.146 	&$	-0.260 	$&$	-1.096 	$&$	-0.312 	$\\
$	^{	205	}	$	Fr	$	\to	^{	201	}$	At	$$&	7.054 	&${9/2^-}\to{(9/2^-)}$&0&	0.112 	&$	0.582 	$&$	-0.370 	$&$	0.507 	$\\
$	^{	207	}	$	Fr	$	\to	^{	203	}$	At	$$&	6.894 	&${9/2^-}\to{9/2^-}$&0&	0.096 	&$	1.190 	$&$	0.171 	$&$	1.167 	$\\
$	^{	209	}	$	Fr	$	\to	^{	205	}$	At	$$&	6.777 	&${9/2^-}\to{9/2^-}$&0&	0.064 	&$	1.753 	$&$	0.561 	$&$	1.719 	$\\
$	^{	211	}	$	Fr	$	\to	^{	207	}$	At	$$&	6.662 	&${9/2^-}\to{9/2^-}$&0&	0.043 	&$	2.328 	$&$	0.958 	$&$	2.379 	$\\
$	^{	213	}	$	Fr	$	\to	^{	209	}$	At	$$&	6.905 	&${9/2^-}\to{9/2^-}$&0&	0.028 	&$	1.535 	$&$	-0.014 	$&$	2.185 	$\\
$	^{	203	}	$	Ra	$	\to	^{	199	}$	Rn	$$&	7.735 	&${(3/2^-)}\to{(3/2^-)}$&0&	0.185 	&$	-1.444 	$&$	-2.177 	$&$	-1.509 	$\\
$	^{	209	}	$	Ra	$	\to	^{	205	}$	Rn	$$&	7.143 	&${5/2^-}\to{5/2^-}$&0&	0.092 	&$	0.673 	$&$	-0.363 	$&$	0.632 	$\\
$	^{	205	}	$Ac		$	\to	^{	201	}$	Fr	$$&	8.096 	&${9/2^-\#}\to{(9/2^-)}$&0&	0.016 	&$	-1.097 	$&$	-2.904 	$&$	-2.261 	$\\
$	^{	207	}	$Ac		$	\to	^{	203	}$	Fr	$$&	7.849 	&${9/2^-\#}\to{9/2^-}$&0&	0.197 	&$	-1.509 	$&$	-2.214 	$&$	-1.477 	$\\
$	^{	211	}	$	Ac	$	\to	^{	207	}$	Fr	$$&	7.619 	&${9/2^-}\to{9/2^-}$&0&	0.117 	&$	-0.672 	$&$	-1.605 	$&$	-0.582 	$\\
$	^{	213	}	$Pa		$	\to	^{	209	}$	Ac	$$&	8.395 	&${9/2^-\#}\to{(9/2^-)}$&0&	0.094 	&$	-2.155 	$&$	-3.184 	$&$	-2.263 	$\\
$	^{	215	}	$Pa		$	\to	^{	211	}$Ac	$$&	8.235 	&${9/2^-\#}\to{9/2^-}$&0&	0.116 	&$	-1.854 	$&$	-2.791 	$&$	-1.591 	$\\
 \noalign{\global\arrayrulewidth1pt}\noalign{\global\arrayrulewidth0.4pt} \multicolumn{8}{c}{\textbf{Nuclei in Regions II and III}}\\
$	^{	177	}	$	Tl	$	\to	^{	173	}$	Au	$$&	7.066 	&${(1/2^+)}\to{(1/2^+)}$&0&	0.223 	&$	-1.609 	$&$	-0.652 	$&$	-1.821 	$\\
$	^{	179	}	$	Tl	$	\to	^{	175	}$	Au	$$&	6.705 	&${1/2^+}\to{1/2^+}$&0&	0.188 	&$	-0.356 	$&$	-0.726 	$&$	-0.634 	$\\
$	^{	213	}	$	Po	$	\to	^{	209	}$	Pb	$$&	8.536 	&${9/2^+}\to{9/2^+}$&0&	0.177 	&$	-5.431 	$&$	-0.752 	$&$	-5.641 	$\\
$	^{	215	}	$	Po	$	\to	^{	211	}$	Pb	$$&	7.527 	&${9/2^+}\to{9/2^+}$&0&	0.193 	&$	-2.750 	$&$	-0.714 	$&$	-2.942 	$\\
$	^{	219	}	$	Po	$	\to	^{	215	}$	Pb	$$&	5.916 	&${9/2^+\#}\to{9/2^+\#}$&0&	0.179 	&$	3.340 	$&$	-0.747 	$&$	3.082 	$\\
$	^{	213	}	$	At	$	\to	^{	209	}$	Bi	$$&	9.254 	&${9/2^-}\to{9/2^-}$&0&	0.274 	&$	-6.903 	$&$	-0.562 	$&$	-6.925 	$\\
$	^{	215	}	$	At	$	\to	^{	211	}$	Bi	$$&	8.178 	&${9/2^-}\to{9/2^-}$&0&	0.112 	&$	-4.000 	$&$	-0.951 	$&$	-4.438 	$\\
$	^{	217	}	$	At	$	\to	^{	213	}$	Bi	$$&	7.202 	&${9/2^-}\to{9/2^-}$&0&	0.273 	&$	-1.487 	$&$	-0.564 	$&$	-1.564 	$\\
$	^{	219	}	$	At	$	\to	^{	215	}$	Bi	$$&	6.342 	&${(9/2^-)}\to{(9/2^-)}$&0&	0.242 	&$	1.777 	$&$	-0.616 	$&$	1.623 	$\\
$	^{	215	}	$	Rn	$	\to	^{	211	}$	Po	$$&	8.839 	&${9/2^+}\to{9/2^+}$&0&	0.249 	&$	-5.638 	$&$	-0.604 	$&$	-5.728 	$\\
$	^{	217	}	$	Rn	$	\to	^{	213	}$	Po	$$&	7.888 	&${9/2^+}\to{9/2^+}$&0&	0.296 	&$	-3.268 	$&$	-0.529 	$&$	-3.316 	$\\
$	^{	215	}	$	Fr	$	\to	^{	211	}$	At	$$&	9.541 	&${9/2^-}\to{9/2^-}$&0&	0.405 	&$	-7.066 	$&$	-0.393 	$&$	-6.937 	$\\
$	^{	217	}	$	Fr	$	\to	^{	213	}$	At	$$&	8.470 	&${9/2^-}\to{9/2^-}$&0&	0.572 	&$	-4.775 	$&$	-0.243 	$&$	-4.538 	$\\
$	^{	219	}	$	Fr	$	\to	^{	215	}$	At	$$&	7.449 	&${9/2^-}\to{9/2^-}$&0&	0.395 	&$	-1.699 	$&$	-0.403 	$&$	-1.662 	$\\
$	^{	217	}	$	Ra	$	\to	^{	213	}$	Rn	$$&	9.161 	&${(9/2^+)}\to{9/2^+\#}$&0&	0.286 	&$	-5.788 	$&$	-0.544 	$&$	-5.845 	$\\
$	^{	215	}	$	Ac	$	\to	^{	211	}$	Fr	$$&	7.746 	&${9/2^-}\to{9/2^-}$&0&	0.046 	&$	-0.770 	$&$	-1.337 	$&$	-1.536 	$\\
$	^{	217	}	$	Ac	$	\to	^{	213	}$	Fr	$$&	9.832 	&${9/2^-}\to{9/2^-}$&0&	0.534 	&$	-7.161 	$&$	-0.272 	$&$	-6.929 	$\\
$	^{	219	}	$	Ac	$	\to	^{	215	}$	Fr	$$&	8.827 	&${9/2^-}\to{9/2^-}$&0&	0.480 	&$	-4.928 	$&$	-0.319 	$&$	-4.800 	$\\
$	^{	219	}	$	Th	$	\to	^{	215	}$	Ra	$$&	9.511 	&${9/2^+\#}\to{9/2^+\#}$&0&	0.312 	&$	-5.991 	$&$	-0.506 	$&$	-6.034 	$\\
$	^{	217	}	$	Pa	$	\to	^{	213	}$	Ac	$$&	8.488 	&${9/2^-\#}\to{9/2^-\#}$&0&	0.082 	&$	-2.458 	$&$	-1.086 	$&$	-2.975 	$\\
$	^{	219	}	$	Pa	$	\to	^{	215	}$	Ac	$$&	10.084 	&${9/2^-}\to{9/2^-}$&0&	0.826 	&$	-7.276 	$&$	-0.083 	$&$	-6.871 	$\\
$	^{	221	}	$	Pa	$	\to	^{	217	}$	Ac	$$&	9.251 	&${9/2^-}\to{9/2^-}$&0&	0.415 	&$	-5.229 	$&$	-0.382 	$&$	-5.193 	$\\
$	^{	221	}	$	U	$	\to	^{	217	}$	Th	$$&	9.889 	&${(9/2^+)}\to{9/2^+\#}$&0&	0.312 	&$	-6.180 	$&$	-0.506 	$&$	-6.246 	$\\
\hline
\end{tabular}
\end{center}

\begin{center}
\tabcaption{Same as Tables\, \ref{table 1} and \ref{table 2}, but for unfavored $\alpha$ decay of odd-$\emph{A}$ nuclei around the doubly magic core at $Z$ = 82, $N$ = 126.}
\label {table 3}
\footnotesize
\begin{tabular}{cccccccc}
\hline {$\mathcal{\alpha}$ transition} & $Q_{\alpha}$ (MeV) & ${j^{\pi}_{p}}\to{j^{\pi}_{d}}$ &$l_{\rm{min}}$ &${P_{\alpha}}$ & $\rm{lg}T^{\rm{expt}}_{1/2}$ (s)&${\rm{lg}T_{1/2}^{\rm{calc1}}}$ (s)& ${\rm{lg}T_{1/2}^{\rm{calc2}}}$ (s)\\ \hline
\noalign{\global\arrayrulewidth1pt}\noalign{\global\arrayrulewidth0.4pt} \multicolumn{8}{c}{\textbf{Nuclei in Region I}}\\
$	^{	209	}	$	Bi	$	\to	^{	205	}$	Tl	$$&	3.138 	&${9/2^-}\to{1/2^+}$ &5&	0.001 	&$	26.802 	$&$	23.934 	$&$	25.999 	$\\
$	^{	189	}	$	Po	$	\to	^{	185	}$	Pb	$$&	7.694 	&${(5/2^-)}\to{3/2^-}$ &2&	0.158 	&$	-2.420 	$&$	-3.223 	$&$	-2.383 	$\\
$	^{	203	}	$	Po	$	\to	^{	199	}$	Pb	$$&	5.496 	&${5/2^-}\to{3/2^-}$ &2&	0.062 	&$	6.294 	$&$	5.085 	$&$	6.353 	$\\
$	^{	205	}	$	Rn	$	\to	^{	201	}$	Po	$$&	6.386 	&${5/2^-\#}\to{3/2^-\#}$ &2&	0.100 	&$	2.837 	$&$	1.840 	$&$	2.843 	$\\
$	^{	207	}	$	Ra	$	\to	^{	203	}$	Rn	$$&	7.269 	&${1/2^-}\to{5/2^-}$ &2&	0.129 	&$	0.205 	$&$	-0.683 	$&$	0.157 	$\\
$	^{	213	}	$	Ra	$	\to	^{	209	}$	Rn	$$&	6.862 	&${(1/2^-)}\to{5/2^(-)}$&2&	0.022 	&$	2.309 	$&$	0.641 	$&$	2.194 	$\\
$	^{	215	}	$	Th	$	\to	^{	211	}$	Ra	$$&	7.665 	&${1/2^-\#}\to{5/2^-\#}$ &2&	0.034 	&$	0.079 	$&$	-1.387 	$&$	0.074 	$\\
\noalign{\global\arrayrulewidth1pt}\noalign{\global\arrayrulewidth0.4pt} \multicolumn{8}{c}{\textbf{Nuclei in Regions II and III}}\\												$^{	187	}	$	Pb	$	\to	^{	183	}$	Hg	$$&	6.393 	&${3/2^-}\to{1/2^-}$ &2&	0.016 	&$	2.203 	$&$	0.406 	$&$	2.471 	$\\
$	^{	189	}	$	Pb	$	\to	^{	185	}$	Hg	$$&	5.915 	&${3/2^-}\to{1/2^-}$ &2&	0.024 	&$	3.989 	$&$	2.375 	$&$	4.440 	$\\
$	^{	213	}	$	Bi	$	\to	^{	209	}$	Ti	$$&	5.988 	&${9/2^-}\to{1/2^+}$ &5&	0.001 	&$	5.116 	$&$	2.256 	$&$	5.214 	$\\
$	^{	213	}	$	Rn	$	\to	^{	209	}$	Po	$$&	8.245 	&${9/2^+\#}\to{1/2^-}$ &5&	0.002 	&$	-1.710 	$&$	-4.444 	$&$	-1.486 	$\\
$	^{	219	}	$	Rn	$	\to	^{	215	}$	Po	$$&	6.946 	&${5/2^+}\to{9/2^+}$ &2&	0.048 	&$	0.598 	$&$	-0.720 	$&$	0.794 	$\\
$	^{	221	}	$	Rn	$	\to	^{	217	}$	Po	$$&	6.162 	&${7/2^+}\to{(9/2^+)}$ &2&	0.040 	&$	3.844 	$&$	2.442 	$&$	3.835 	$\\
$	^{	215	}	$	Ra	$	\to	^{	211	}$	Rn	$$&	8.864 	&${9/2^+\#}\to{1/2^-}$ &5&	0.003 	&$	-2.777 	$&$	-5.364 	$&$	-2.916 	$\\
$	^{	219	}	$	Ra	$	\to	^{	215	}$	Rn	$$&	8.138 	&${(7/2^+)}\to{9/2^+}$ &2&	0.018 	&$	-2.000 	$&$	-3.754 	$&$	-2.274 	$\\
$	^{	217	}	$	Th	$	\to	^{	213	}$	Ra	$$&	9.435 	&${9/2^+\#}\to{1/2^-}$ &5&	0.004 	&$	-3.607 	$&$	-6.048 	$&$	-3.828 	$\\
\hline
\end{tabular}
\end{center}

\begin{center}
\tabcaption{Same as Tables\, \ref{table 1} and \ref{table 2}, but for favored $\alpha$ decay of doubly-odd nuclei.}
\label {table 4}
\footnotesize
\begin{tabular}{cccccccc}
\hline {$\mathcal{\alpha}$ transition} & $Q_{\alpha}$ (MeV) & ${j^{\pi}_{p}}\to{j^{\pi}_{d}}$ &$l_{\rm{min}}$ &${P_{\alpha}}$ & $\rm{lg}T^{\rm{expt}}_{1/2}$ (s)&${\rm{lg}T_{1/2}^{\rm{calc1}}}$ (s)& ${\rm{lg}T_{1/2}^{\rm{calc2}}}$ (s)\\ \hline
\noalign{\global\arrayrulewidth1pt}\noalign{\global\arrayrulewidth0.4pt} \multicolumn{8}{c}{\textbf{Nuclei in Region I}}\\
$	^{	192	}	$	At	$	\to	^{	188	}$	Bi	$$&	7.696 	&${3^+\#}\to{3^+\#}$ &0&	0.115 	&$	-1.939$&$	-2.957 	$&$	-2.019 	$\\
$	^{	200	}	$	At	$	\to	^{	196	}$	Bi	$$&	6.596 	&${(3^+)}\to{(3^+)}$ &0&	0.059 	&$	1.917	$&$	0.594 	$&$	1.823 	$\\
$	^{	202	}	$	At	$	\to	^{	198	}$	Bi	$$&	6.353 	&${3^(+)}\to{3^(+)}$ &0&	0.045 	&$	3.161	$&$	1.511 	$&$	2.857 	$\\
$	^{	204	}	$	At	$	\to	^{	200	}$	Bi	$$&	6.071 	&${7^+}\to{7^+}$ &0&	0.031 	&$	4.156$&$	2.689 	$&$	4.198 	$\\
$	^{	206	}	$	At	$	\to	^{	202	}$	Bi	$$&	5.886 	&${(5)^+}\to{5^(+\#)}$ &0&	0.017 	&$	5.306	$&$	3.506 	$&$	5.279 	$\\
$	^{	208	}	$	At	$	\to	^{	204	}$	Bi	$$&	5.751 	&${6^+}\to{6^+}$ &0&	0.003 	&$	6.023$&$	4.114 	$&$	6.667 	$\\
$	^{	200	}	$	Fr	$	\to	^{	196	}$	At	$$&	7.615 	&${(3^+)}\to{(3^+)}$ &0&	0.134 	&$	-1.323	$&$	-2.113 	$&$	-1.240 	$\\
$	^{	204	}	$	Fr	$	\to	^{	200	}$	At	$$&	7.170 	&${3^+}\to{(3^+)}$ &0&	0.087 	&$	0.260	$&$	-0.764 	$&$	0.295 	$\\
$	^{	206	}	$	Fr	$	\to	^{	202	}$	At	$$&	6.924 	&${3^+}\to{3^(+)}$ &0&	0.064 	&$	1.258	$&$	0.084 	$&$	1.279 	$\\
$	^{	208	}	$	Fr	$	\to	^{	204	}$	At	$$&	6.784 	&${7^+}\to{7^+}$ &0&	0.040 	&$	1.821	$&$	0.555 	$&$	1.950 	$\\
$	^{	206	}	$	Ac	$	\to	^{	202	}$	Fr	$$&	7.959 	&${(3^+)}\to{3^+}$ &0&	0.129 	&$	-1.602	$&$	-2.528 	$&$	-1.640 	$\\
\noalign{\global\arrayrulewidth1pt}\noalign{\global\arrayrulewidth0.4pt} \multicolumn{8}{c}{\textbf{Nuclei in Regions II and III}}\\										$	^{	214	}	$	At	$	\to	^{	210	}$	Bi	$$&	8.987 	&${1^-}\to{1^-}$ &0&	0.186 	&$	-6.253	$&$	-6.912 	$&$	-6.182 	$\\
$	^{	216	}	$	At	$	\to	^{	212	}$	Bi	$$&	7.950 	&${1^(-)}\to{1^(-)}$ &0&	0.218 	&$	-3.523	$&$	-4.336 	$&$	-3.674 	$\\
$	^{	218	}	$	At	$	\to	^{	214	}$	Bi	$$&	6.874 	&${1^-\#}\to{1^-}$ &0&	0.249 	&$	0.176$&$	-0.910 	$&$	-0.307 	$\\
$	^{	216	}	$	Fr	$	\to	^{	212	}$	At	$$&	9.175 	&${(1^-)}\to{(1^-)}$ &0&	0.218 	&$	-6.155	$&$	-6.685 	$&$	-6.022 	$\\
$	^{	218	}	$	Fr	$	\to	^{	214	}$	At	$$&	8.014 	&${1^-}\to{1^-}$ &0&	0.270 	&$	-3.000	$&$	-3.791 	$&$	-3.223 	$\\
$	^{	218	}	$	Ac	$	\to	^{	214	}$	Fr	$$&	9.374 	&${1^-\#}\to{(1^-)}$ &0&	0.249 	&$	-6.000	$&$	-6.487 	$&$	-5.884 	$\\
$	^{	220	}	$	Pa	$	\to	^{	216	}$	Ac	$$&	9.651 	&${1^-\#}\to{(1^-)}$ &0&	0.281 	&$	-6.108	$&$	-6.471 	$&$	-5.920 	$\\
\hline
\end{tabular}
\end{center}

\begin{center}
\tabcaption{Same as Tables\, \ref{table 1} and \ref{table 2}, but for unfavored $\alpha$ decay of doubly-odd nuclei.}
\label {table 5}
\footnotesize
\begin{tabular}{cccccccc}
\hline {$\mathcal{\alpha}$ transition} & $Q_{\alpha}$ (MeV) & ${j^{\pi}_{p}}\to{j^{\pi}_{d}}$ &$l_{\rm{min}}$ &${P_{\alpha}}$ & $\rm{lg}T^{\rm{expt}}_{1/2}$ (s)&${\rm{lg}T_{1/2}^{\rm{calc1}}}$ (s)& ${\rm{lg}T_{1/2}^{\rm{calc2}}}$ (s)\\ \hline
\noalign{\global\arrayrulewidth1pt}\noalign{\global\arrayrulewidth0.4pt} \multicolumn{8}{c}{\textbf{Nuclei in Region I}}\\
$	^{	186	}	$	Bi	$	\to	^{	182	}$	Tl	$$&	7.757 	&${(3^+)}\to{(2^-)}$ &1&	0.012 	&$	-1.830 	$&$	-3.749 	$&$	-2.092 	$\\
$	^{	190	}	$	Bi	$	\to	^{	186	}$	Tl	$$&	6.862 	&${(3^+)}\to{(2^-)}$ &1&	0.012 	&$	0.912 	$&$	0.688 	$&$	3.704 	$\\
$	^{	192	}	$	Bi	$	\to	^{	188	}$	Tl	$$&	6.381 	&${(3^+)}\to{(2^-)}$ &1&	0.020 	&$	2.442 	$&$	2.478 	$&$	2.852 	$\\
$	^{	194	}	$	Bi	$	\to	^{	190	}$	Tl	$$&	5.918 	&${(3^+)}\to{(2^-)}$ &1&	0.023 	&$	4.313 	$&$	4.420 	$&$	2.697 	$\\
$	^{	210	}	$	At	$	\to	^{	206	}$	Bi	$$&	5.631 	&${(5)^+}\to{(6)^+}$ &2&	0.003 	&$	7.221 	$&$	6.687 	$&$	6.125 	$\\
$	^{	210	}	$	Fr	$	\to	^{	206	}$	At	$$&	6.672 	&${6^+}\to{(5)^+}$ &2&	0.038 	&$	2.427 	$&$	2.755 	$&$	2.019 	$\\
$	^{	212	}	$	Fr	$	\to	^{	208	}$	At	$$&	6.529 	&${5^+}\to{6^+}$ &2&	0.012 	&$	3.444 	$&$	3.442 	$&$	3.626 	$\\
$	^{	212	}	$	Pa	$	\to	^{	208	}$	Ac	$$&	8.415 	&${7^+\#}\to{(3^+)}$ &4&	0.021 	&$	-2.125 	$&$	-3.794 	$&$	-2.116 	$\\
\noalign{\global\arrayrulewidth1pt}\noalign{\global\arrayrulewidth0.4pt} \multicolumn{8}{c}{\textbf{Nuclei in Regions II and III}}\\										
$	^{	210	}	$	Bi	$	\to	^{	206	}$	Tl	$$&	5.037 	&${1^-}\to{0^-}$ &2&	2.330E-05	&$	11.616 	$&$	10.343 	$&$	21.455 	$\\
$	^{	212	}	$	Bi	$	\to	^{	208	}$	Tl	$$&	6.207 	&${1^(-)}\to{5^+}$ &5&	0.002 	&$	4.005 	$&$	4.171 	$&$	7.326 	$\\
$	^{	214	}	$	Bi	$	\to	^{	210	}$	Tl	$$&	5.621 	&${1^-}\to{5^+\#}$ &5&	0.002 	&$	6.753 	$&$	6.667 	$&$	7.478 	$\\
$	^{	212	}	$	At	$	\to	^{	208	}$	Bi	$$&	7.817 	&${(1^-)}\to{5^+}$ &5&	0.001 	&$	-0.503 	$&$	-0.719 	$&$	9.542 	$\\
$	^{	214	}	$	Fr	$	\to	^{	210	}$	At	$$&	8.588 	&${(1^-)}\to{(5)^+}$ &5&	0.002 	&$	-2.286 	$&$	-2.359 	$&$	7.413 	$\\
$	^{	216	}	$	Ac	$	\to	^{	212	}$	Fr	$$&	9.235 	&${(1^-)}\to{5^+}$ &5&	0.003 	&$	-3.357 	$&$	-3.423 	$&$	6.596 	$\\
\hline
\end{tabular}
\end{center}

\begin{multicols}{2}

\begin{center}
\tabcaption{The fitted parameters of Eq.\,(\ref{subeq:16}).}
\label {table 6}
\footnotesize
\begin{tabular}{ccccc}
\hline
\multicolumn{1}{c}{\multirow{2}{*}{Region}}&\multicolumn{2}{c}{\text{favored decay}}&\multicolumn{2}{c}{\text{unfavored decay}}\\
\cline{2-3}  \cline{4-5}
 &a&b&a&b\\
\hline
\noalign{\global\arrayrulewidth1pt}\noalign{\global\arrayrulewidth0.4pt}&\multicolumn{4}{c}{\textbf{even-even Nuclei}}\\
I&-0.66547&0.03339&{-}&{-}\\
II,III&1.8334&0.25035&{-}&{-}\\
\noalign{\global\arrayrulewidth1pt}\noalign{\global\arrayrulewidth0.4pt} &\multicolumn{4}{c}{\textbf{odd-$A$ Nuclei}}\\
I&-0.65688&0.00632&-0.67342&0.00862\\
II,III&0.64805&0.26947&0.2559&-0.00382\\
\noalign{\global\arrayrulewidth1pt}\noalign{\global\arrayrulewidth0.4pt}& \multicolumn{4}{c}{\textbf{doubly-odd Nuclei}}\\
I&-0.48781&-0.01831&-0.10994&0.00988\\
II,III&1.0972&0.13849&0.09388&$-1.34923\times10^{-4}$\\
\hline
\end{tabular}
\end{center}

\begin{center}
\tabcaption{ The standard deviations $\sigma$ between the calculated $\alpha$ decay half-lives and experimental ones.}
\label {table 7}
\footnotesize
\begin{tabular}{ccccc}
\hline
\multicolumn{1}{c}{\multirow{2}{*}{Nuclei}}&\multicolumn{2}{c}{\text{favored decay}}&\multicolumn{2}{c}{\text{unfavored decay}}\\
 \cline{2-3}  \cline{4-5}
 &$\sigma_1$&$\sigma_2$&$\sigma_1$&$\sigma_2$\\
\hline
Even-even nuclei&0.947&0.350&{-}&{-}\\
$Odd-A$ nuclei &0.978&0.262&1.90&0.272\\
Doubly odd nuclei &1.145&0.227&2.494&0.400\\
\hline
\end{tabular}
\end{center}

\begin{center}
\includegraphics[width=9cm]{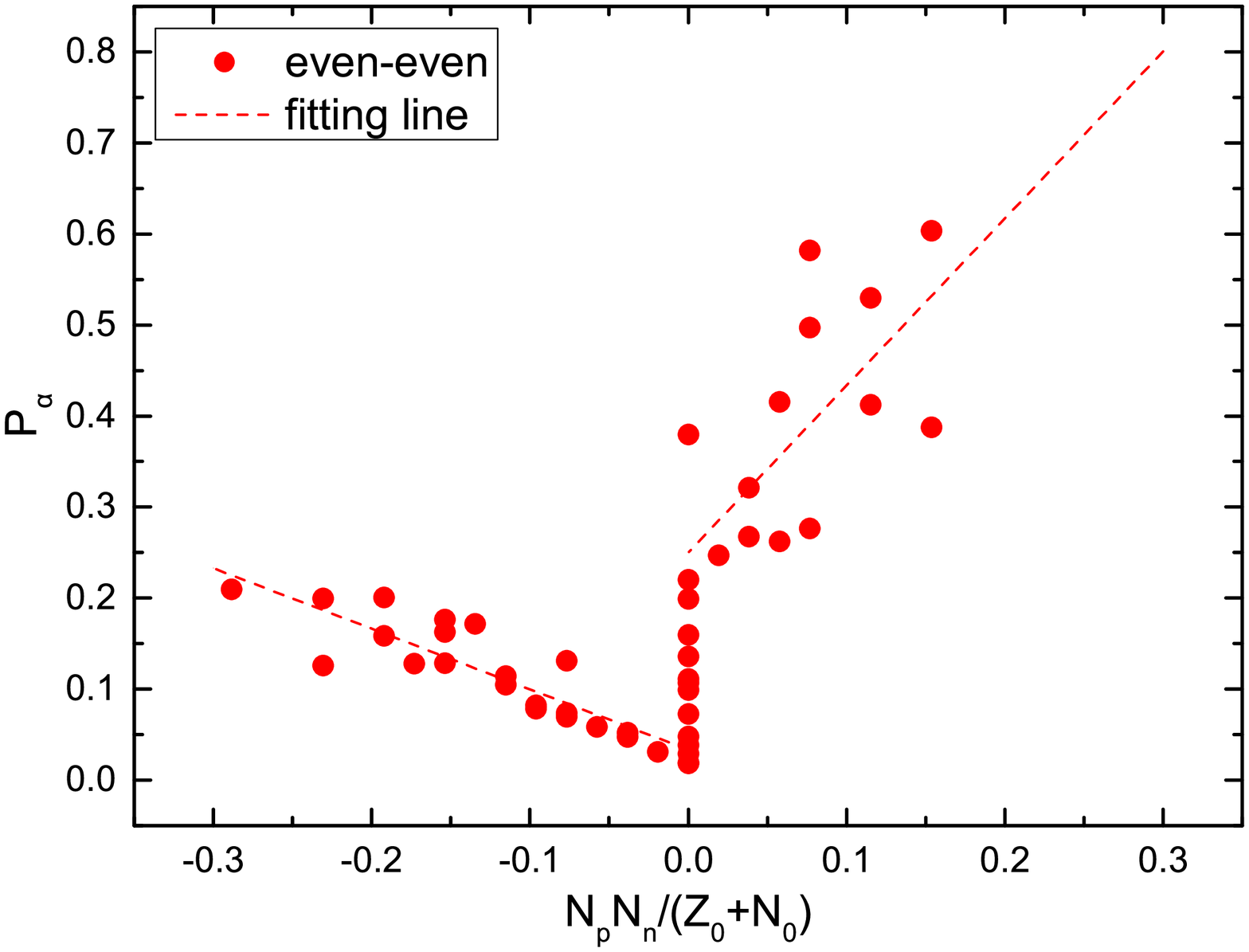}
\figcaption{\label{figure1}(color online) The $\alpha$ preformation factors of even-even nuclei around $Z_0$ = 82 and $N_0$ = 126 shell closures as a function of $\frac{N_pN_n}{N_0 + Z_0}$, where $N_p$ and $N_n$ denote valence protons (holes) and neutrons (holes) of parent nucleus, respectively. The dash lines are the fittings of $\alpha$ preformation factors.}
\label {fig 1}
\end{center}

\begin{center}
\includegraphics[width=9cm]{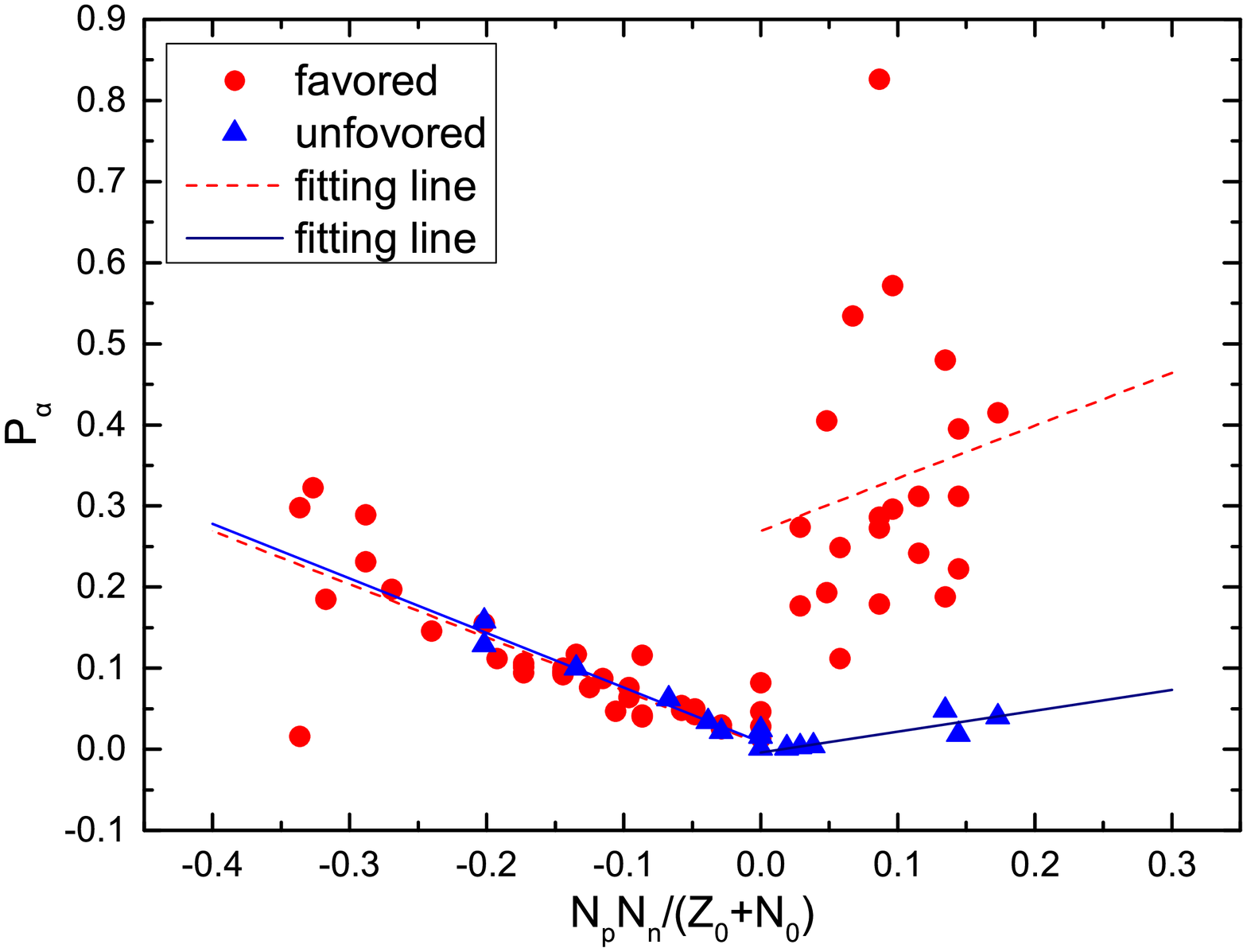}
\figcaption{\label{figure2} (color online) Same as Fig.\,\ref{fig 1}, but it represents the $\alpha$ preformation factors as a function of $\frac{N_pN_n}{N_0 + Z_0}$ of odd-$\emph{A}$ nuclei.}
\label {fig 2}
\end{center}

\begin{center}
\includegraphics[width=9cm]{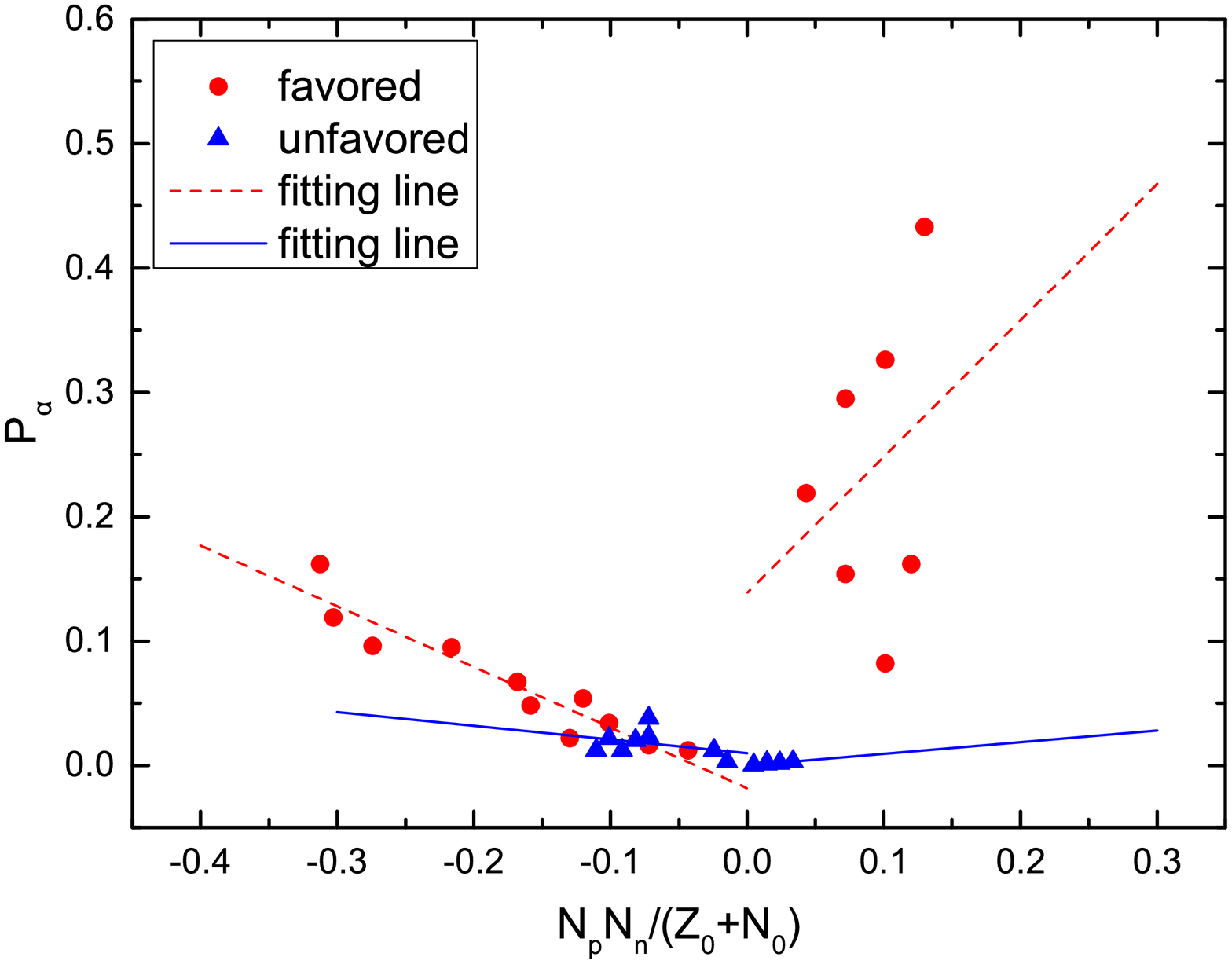}
\figcaption{\label{figure3} (color online) Same as Fig.\,\ref{fig 1}, but it represents the $\alpha$ preformation factors as a function of $\frac{N_pN_n}{N_0 + Z_0}$ of doubly-odd nuclei.}
\label {fig 3}
\end{center}

\begin{center}
\includegraphics[width=9cm]{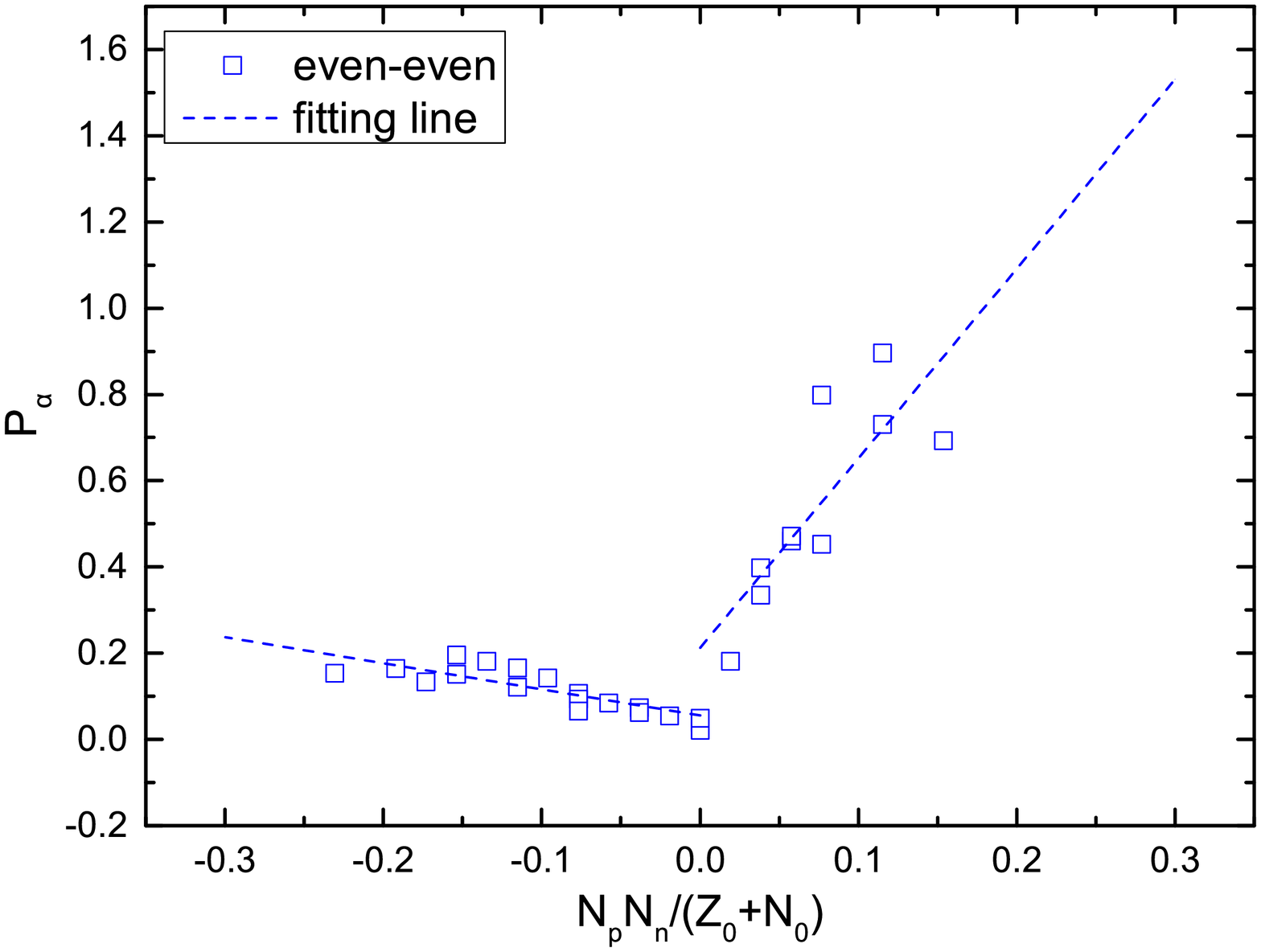}
\figcaption{\label{figure2} (color online) The $\alpha$ preformation factors $P_{\alpha}$ of even-even nuclei around $Z_0$ = 82 and $N_0$ = 126 shell closures as a function of $\frac{N_pN_n}{N_0 + Z_0}$, which $P_{\alpha}$ are obtained by eq.\,(\ref{subeq:14}). The blue opened squares denote the even-even nuclei around shell closures, the blue dashed lines are the fittings of the $P_{\alpha}$.}
\label {fig 4}
\end{center}

\begin{center}
\includegraphics[width=9cm]{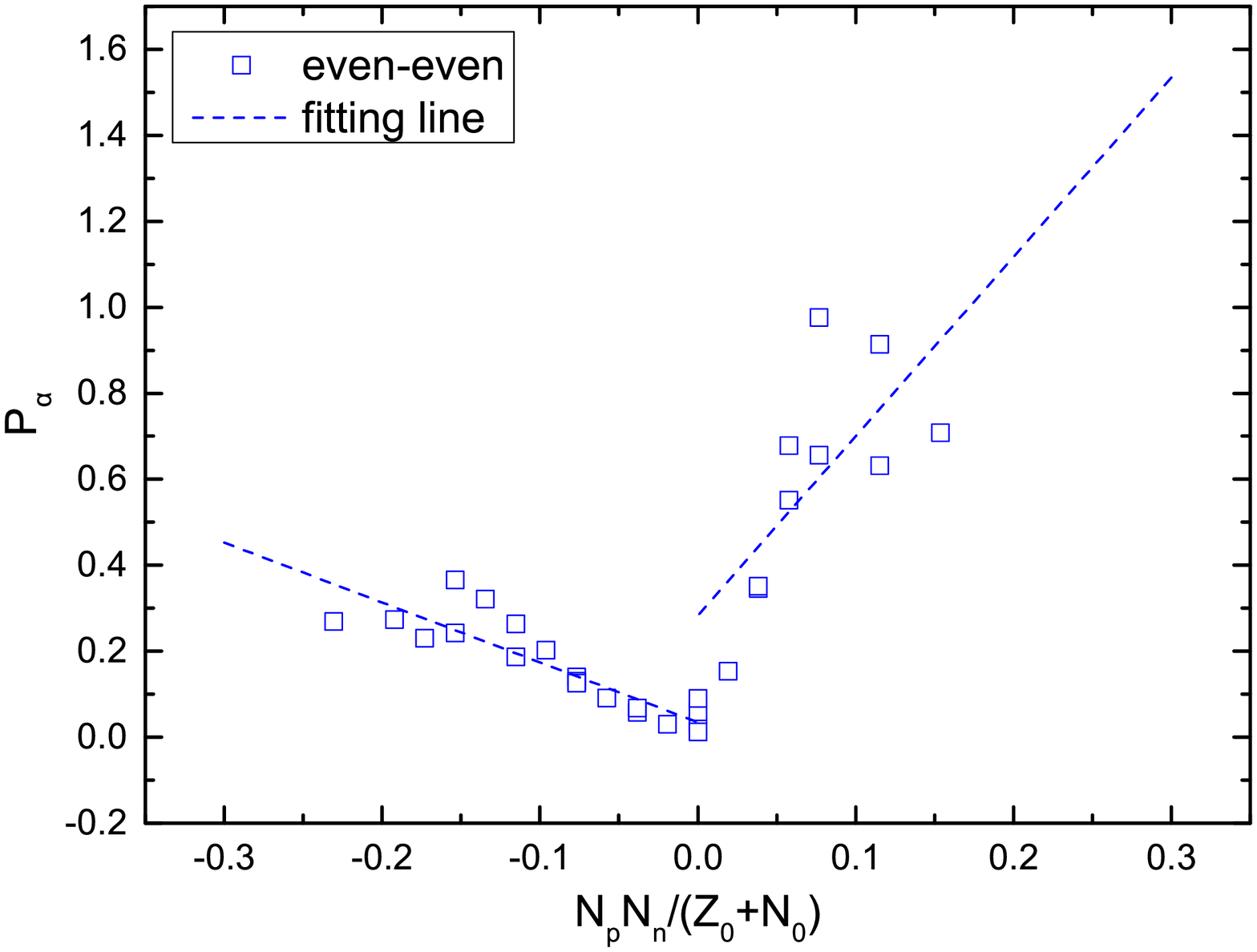}
\figcaption{\label{figure3} (color online) Same as Fig.\,\ref{fig 4}, but the $\alpha$ preformation factors $P_{\alpha}$ are calculated by eq.\,(\ref{subeq:15}).}
\label {fig 5}
\end{center}

\end{multicols}

\begin{multicols}{2}

\section{Summary}

In summary, using the generalized liquid drop model (GLDM), we systematically study the $\alpha$ decay half-lives and the preformation factors of 152 nuclei around $\emph{Z}$ = 82, $\emph{N}$ = 126 closed shells. It is found that the preformation factors are linearly related to $N_pN_n$ and the calculated half-lives can well reproduce the experimental data. Meanwhile, the linear relationship between  $N_pN_n$ and the preformation factors calculated by two different formulas by defining the concept of the microscopic valence nucleon (hole) number still exists. Combining with Seif ${et\ al.}$ work and our previous works, we think that the linear relationship between $P_\alpha$ and $N_pN_n$ around $\emph{Z}$ = 82, $\emph{N}$ = 126 shell closures is not model dependent and the valance proton-neutron interaction may be important to the $\alpha$ particle preformation.
\end{multicols}
\vspace{-1mm}
\centerline{\rule{80mm}{0.1pt}}
\vspace{2mm}

\begin{multicols}{2}

\end{multicols}

\clearpage
\end{CJK*}
\end{document}